\documentclass[journal,transmag]{IEEEtran}

\usepackage{amsmath}
\usepackage{amssymb}
\usepackage{upgreek}
\usepackage{graphicx}
\usepackage{multirow}
\usepackage[caption=false]{subfig}
\usepackage{xcolor}
\usepackage{dsfont}
\usepackage{nidanfloat}
\graphicspath{{figures/}}

\begin{document}

\title{Flow Field Tomography with Uncertainty Quantification using a Bayesian Physics-Informed Neural Network}

\author{\IEEEauthorblockN{Joseph P. Molnar and Samuel J. Grauer}
\IEEEauthorblockA{Department of Mechanical Engineering,
Pennsylvania State University, University Park, PA 16802 USA}
\thanks{Corresponding author: S.J. Grauer (email: sgrauer@psu.edu).}}

\IEEEtitleabstractindextext{
\begin{abstract}
We report a new approach to flow field tomography that uses the Navier--Stokes and advection--diffusion equations to regularize reconstructions. Tomography is increasingly employed to infer 2D or 3D fluid flow and combustion structures from a series of line-of-sight (LoS) integrated measurements using a wide array of imaging modalities. The high-dimensional flow field is reconstructed from low-dimensional measurements by inverting a projection model that comprises path integrals along each LoS through the region of interest. Regularization techniques are needed to obtain realistic estimates, but current methods rely on truncating an iterative solution or adding a penalty term that is incompatible with the flow physics to varying degrees. Physics-informed neural networks (PINNs) are new tools for inverse analysis that enable regularization of the flow field estimates using the governing physics. We demonstrate how a PINN can be leveraged to reconstruct a 2D flow field from sparse LoS-integrated measurements with no knowledge of the boundary conditions by incorporating the measurement model into the loss function used to train the network. The resulting reconstructions are remarkably superior to reconstructions produced by state-of-the-art algorithms, even when a PINN is used for post-processing. However, as with conventional iterative algorithms, our approach is susceptible to semi-convergence when there is a high level of noise. We address this issue through the use of a Bayesian PINN, which facilitates comprehensive uncertainty quantification of the reconstructions, enables the use of a more intuitive loss function, and reveals the source of semi-convergence.
\end{abstract}

\begin{IEEEkeywords}
Bayesian inference, fluid diagnostics, inverse problems, physics-informed neural networks, regularization, tomography, uncertainty quantification
\end{IEEEkeywords}}

\maketitle
\IEEEdisplaynontitleabstractindextext
\IEEEpeerreviewmaketitle

\section{Introduction}
\label{sec:intro}
\IEEEPARstart{N}{on-intrusive}, spatially-resolved measurement techniques are required for research on fluid dynamics in order to observe novel phenomena, develop models of flow behavior, and provide data for the validation of computational codes~\cite{Goldstein2017}. Quantitative optical diagnostics can be used to probe a target flow without disrupting its development, but many optical measurements are confined to a cross-section of the flow field or provide line-of-sight (LoS) integrated information and thereby fail to fully resolve the variables of interest. Flow field tomography employs multiple simultaneous LoS measurements, called projections, in conjunction with a reconstruction algorithm to infer quantitative 2D or 3D distributions of key variables (velocity, temperature, mole fractions, etc.). This paper describes a new framework for tomographic imaging of flow fields that is based on physics-informed neural networks (PINNs)~\cite{Raissi2019,Raissi2020}. Our approach significantly improves upon state-of-the-art methods for estimating the quantity (or quantities) of interest (QoI), including signal processing techniques that employ a PINN to post-process volumetric flow field measurements. Furthermore, we use a Bayesian implementation to conduct uncertainty quantification (UQ), which is needed for scientific measurements but is often overlooked in flow field tomography due to complications associated with ill-posed inverse problems~\cite{Kaipio2006}.\par

Numerous tomographic modalities have been developed to characterize reacting and non-reacting flows that contain gas-, liquid-, and/or solid-phase constituents. Common variants include tomographic particle image velocimetry (PIV)~\cite{Scarano2012}, laser absorption spectroscopy~\cite{Cai2017}, chemiluminescence~\cite{Floyd2011}, laser-induced fluorescence~\cite{Wu2015}, laser-induced incandescence~\cite{Meyer2016}, X-ray imaging~\cite{Halls2014}, and background-oriented schlieren~\cite{Grauer2018}, among others. These diagnostics are united by common mathematical properties. In each case, individual projections are described by a path or volume integral through the flow field, and the full set of integrals for a measurement system constitutes a ``forward model.'' Reconstruction consists in inverting this model for a series of measurements to estimate a 2D or 3D distribution of the QoI. However, reconstruction is an ill-posed problem that inherently amplifies noise and is almost always rank deficient in the context of flow field tomography~\cite{Daun2016}. In other words, absent an impracticable quantity of projections or an axisymmetric target, there exists an infinite set of solutions that can fully satisfy the measurements. As a result, regularization is used to incorporate additional information about the flow into the reconstruction procedure to obtain a unique, physically-plausible estimate for each set of projections.\par

In this paper, we use conventional and Bayesian PINNs (C-PINNs and B-PINNs) to \textit{directly} reconstruct flow field variables from a limited set of scalar field projections. This technique is applicable to numerous tomographic modalities, including those mentioned earlier. PINNs can incorporate the equations governing fluid motion through the use of a physics loss term to estimate various QoI, and previous work employed a PINN in the context of flow field tomography to infer velocity fields from the temperature field reconstructed by BOS tomography~\cite{Cai2021a}. We show how a PINN can be used for tomographic reconstruction per se, as opposed to post-processing reconstructions with a PINN, by embedding the projection model into the loss function. Our approach reduces errors associated with the initial reconstruction algorithm, which are unavoidable when using the post-processing technique. Moreover, we show how a B-PINN not only facilitates UQ but can effectively utilize a wider range of loss functions via the prior than can a C-PINN, including loss functions that place a greater emphasis on the (known) flow field physics. The resulting posterior distribution is suitable for model checking and benchmarking of numerical codes.\par

\section{Flow Field Tomography}
\label{sec:tomography}
In general, tomographic measurements can be described by a simple path integral,
\begin{equation}
    b_i = \int_0^L c\left[\mathbf{r}_i(s)\right] \mathrm{d}s \approx \sum_{j=1}^n A_{i,j} c_j,
    \label{equ:model continuous}
\end{equation}
where $b_i$ is a projection (LoS-integrated measurement) from the $i$th LoS, corresponding to a pixel or laser beam; $c$ is the QoI, such as a scalar concentration field; $\mathbf{r}_i$ is an indicator function in $\mathds{R}^2$ or $\mathds{R}^3$ that picks out a position along the $i$th measurement path; and $s$ is a progress variable that takes values from 0 to the path length, $L$. The measurement domain must be discretized, typically using pixels or triangle elements in 2D or voxels in 3D. Next, the equality in \eqref{equ:model continuous} is approximated using this basis; the discrete approximation is shown on the right side of the equation, where $c_j$ is the value of $c$ at the $j$th basis function (pixel, node, voxel) and $A_{i,j}$ is the path (or volume) integral along the $i$th LoS over the $j$th basis function. By convention, the imaging system comprises $m$ lines-of-sight; the basis contains $n$ functions; the data and QoI are arranged as vectors, accordingly, $\mathbf{b} = \{b_i\}_{i=1}^{m}$ and $\mathbf{c} = \{c_j\}_{j=1}^{n}$; and the measurement sensitivities ($A_{i,j}$) for each LoS and basis function are collated into an $m\times n$ matrix, $\mathbf{A}$. As a result, tomographic projections can be modeled by matrix multiplication,
\begin{equation}
    \mathbf{Ac} = \mathbf{b},
    \label{equ:model discrete}
\end{equation}
and reconstruction consists in inferring the unknown distribution $\mathbf{c}$ that explains the measurements in $\mathbf{b}$. Flow field tomography often features time-resolved data. For convenience, the scalar fields and measurements may be stored in matrix form:
\begin{equation}
    \mathbf{C} = [\mathbf{c}^{(1)}, \mathbf{c}^{(2)},\dots, \mathbf{c}^{(K)}]
    \label{equ:QoI matrix}
\end{equation}
and
\begin{equation}
    \mathbf{B} = [\mathbf{b}^{(1)}, \mathbf{b}^{(2)},\dots, \mathbf{b}^{(K)}],
    \label{equ:measurement matrix}
\end{equation}
where $K$ is the total number of measurements.\par

\begin{figure}[t]
    \centering
    \includegraphics[width=.475\textwidth]{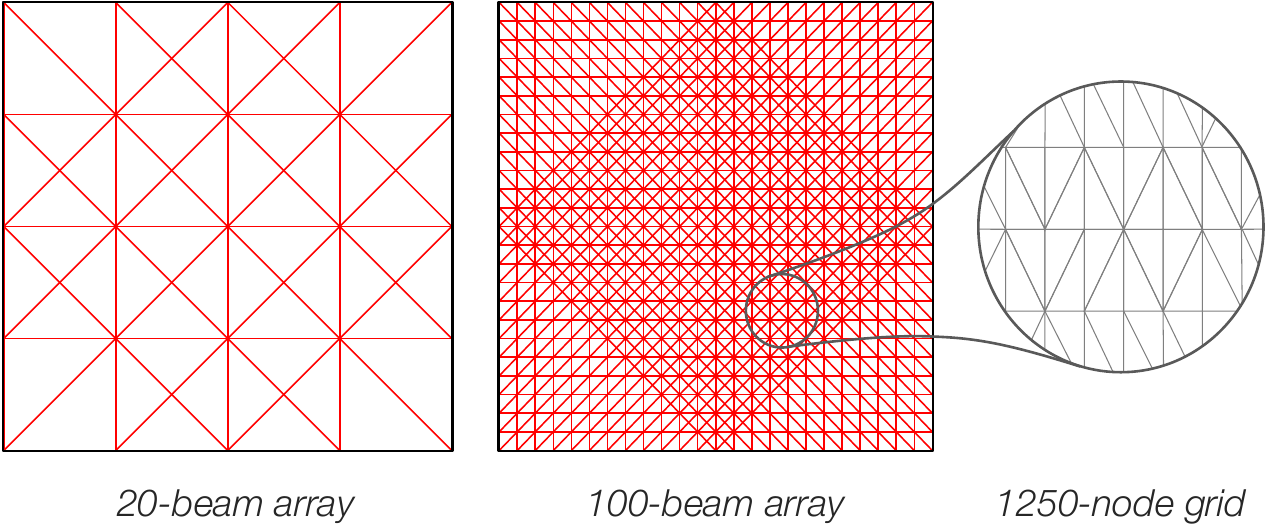}
    \caption{Square domain used in our numerical demonstration of 2D flow field tomography, including a 20-beam and 100-beam array. The zoom bubble depicts a section of the low-resolution, 1250-node finite element mesh used for reconstruction. Not shown are the 40- and 160-beam arrays as well as the 3600-node mesh that were also used in testing.}
    \label{fig:array}
\end{figure}

\subsection{Explicit Reconstruction Algorithms}
\label{sec:tomography:explicit algorithms}
Reconstructing $\mathbf{c}$ from $\mathbf{b}$ is necessarily ill-posed for one of two reasons. Either
\begin{enumerate}
    \item the column rank of $\mathbf{A}$ equals $n$, in which case the compact kernel in \eqref{equ:model continuous} yields an ill-conditioned matrix such that the pseudoinverse of $\mathbf{A}$ amplifies noise in $\mathbf{b}$, discretization errors, calibration errors, and the like, or
    \item there are fewer linearly-independent lines-of-sight than basis functions such that $\mathbf{A}$ has a nontrivial null space and there exists an infinite set of vectors $\mathbf{c}$ that each perfectly solves \eqref{equ:model discrete}~\cite{Daun2016}.
\end{enumerate}
Most imaging systems fall into the latter category. Historically, iterative solvers, such as the additive algebraic reconstruction technique (ART)~\cite{Gordon1970}, multiplicative ART (MART), and simultaneous iterative reconstruction techniques (SIRTs)~\cite{Hansen2018} have been used to solve \eqref{equ:model discrete}. Additive ART and SIRT algorithms approach the least-squares solution or a matrix-weighted Euclidean norm of $\mathbf{c}$ (the matrix norm differs depending on the SIRT variant)~\cite{Hansen2018}, whereas the MART approaches the vector $\mathbf{c}$ that best satisfies \eqref{equ:model discrete} and maximizes the Kullback--Leibler divergence between the initial and converged vectors~\cite{Herman1976}. The latter criterion promotes spatially-sparse solutions when a voxel basis is employed, which has a direct physical analog to particle fields in tomographic PIV. For this reason, derivatives of the MART algorithm remain ubiquitous in tomographic PIV, e.g., \cite{Novara2010}, and are the basis of several major commercial codes.\par

All three classes of iterative techniques are sensitive to noise due to the ill-conditioned nature of $\mathbf{A}$. However, ART, MART, and SIRT algorithms exhibit semi-convergence, meaning that initial iterations contribute low-frequency solution components that are robust to noise~\cite{Elfving2014}. Accordingly, iterative solvers are typically truncated in an ad-hoc manner to promote smooth estimates of the QoI, called \textit{iterative regularization}. Still, the lack of explicit spatial (or temporal) information in these algorithms leads to large reconstruction errors due to the low number of viewing angles that is characteristic of flow field tomography setups.\par

\begin{figure*}[t]
    \centering
    \includegraphics[width=14cm]{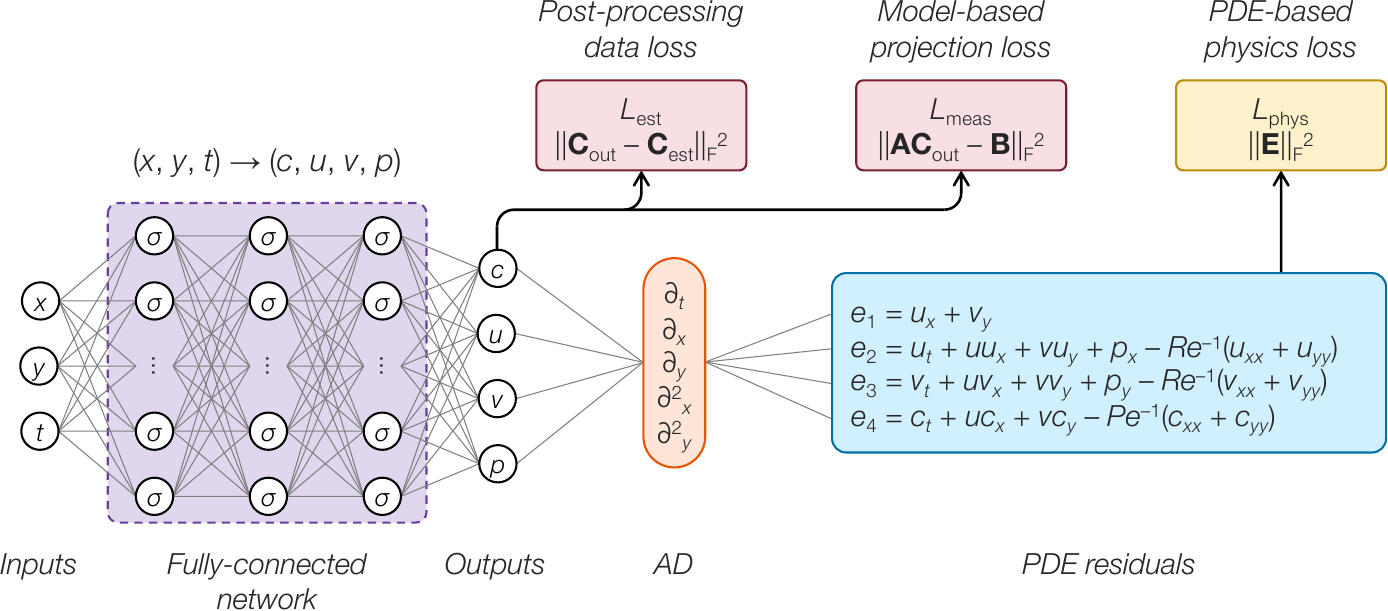}
    \caption{Architecture of a PINN used in 2D flow field tomography: the network has a deep, fully-connected structure that maps spatio-temporal coordinates fed into the input layer, $(x,y,t)$, to QoI, $c, u, v, p$. Partial derivatives of the inputs with respect to the outputs are computed by AD and plugged into the governing PDEs, and the residuals are collected into a physics loss term. The data loss may be constructed in one of two ways: (1) outputted concentrations ($\mathbf{C}_\mathrm{out}$) are compared to concentration data from a conventional reconstruction algorithm ($\mathbf{C}_\mathrm{est}$), which may contain significant errors, or (2) the PINN output is used to predict projection data via the measurement model ($\mathbf{AC}_\mathrm{out}$), and the predicted measurements are compared to experimental projections, $\mathbf{B}$. The former method, \textit{post-processing}, was introduced in \cite{Cai2021a}, and the latter method, \textit{direct reconstruction}, is the focus of this paper.}
    \label{fig:PINN architecture}
\end{figure*}

Numerous explicit regularization schemes have been proposed to overcome the limitations associated with iterative techniques, although iterative methods are often still used to solve an augmented (i.e., regularized) system of equations. Early examples of explicit regularization featured the use of spatial filtering to smooth $\mathbf{c}$ in between each iteration of a SIRT solver~\cite{Terzija2008} or the use of a wavelet basis to control the frequency content of reconstructions~\cite{Terzija2010}. Classical methods of regularization have also been adapted for flow field tomography, including Tikhonov~\cite{Daun2006} and total variation (TV)~\cite{Cai2013} regularization. These methods add a penalty term to the residual of \eqref{equ:model discrete} in order to avoid solutions that have undesirable characteristics. The Tikhonov penalty utilizes the discrete Laplacian of $\mathbf{c}$ to produce spatially-smooth reconstructions, and TV regularization is based on the Manhattan norm of the gradient of $\mathbf{c}$, which also yields smooth estimates but may permit sharp discontinuities between smooth sub-regions of the domain. Bayesian algorithms were initially introduced to flow field tomography to interpret Tikhonov and TV reconstructions for the purpose of UQ~\cite{Kaipio2006,Kolehmainen1998}, but the Bayesian framework also enables the use of explicit statistical information about a flow field through the use of auto-correlation functions~\cite{Grauer2017,Grauer2019}. Each of these techniques has been deployed to process data from a wide variety of 2D and 3D flow field tomography tests. Nevertheless, without exception, the supplemental information is very general (usually advancing some form of smoothness), and the penalty terms are not fully-compatible with the flow physics. Moreover, smoothing techniques intrinsically limit the spatial resolution of a tomographic sensor~\cite{Emmert2020} so an alternative approach is desirable.\par

\subsection{Deep Learning Reconstruction Algorithms}
\label{sec:tomography:deep learning algorithms}
Recent progress in machine learning, brought about by deep neural nets (DNNs), cheap computing power, and large troves of data, has enabled all manner of progress in image classification, natural language processing, and control applications to name a few~\cite{LeCun2015}. Convolutional neural nets (CNNs) have been particularly successful in the realm of image processing and were therefore adapted for tomographic reconstruction by numerous groups. For instance, Huang et al.~\cite{Huang2018,Huang2019,Huang2020} conducted 2D laser absorption tomography and 3D chemiluminescence tomography via CNNs, and Wei et al.~\cite{Wei2020,Wei2021} used CNNs to reconstruct 3D mole fraction and temperature fields from laser absorption measurements of methane and ethylene flame doublets. These examples employed the common supervised training paradigm, in which phantom distributions ($\mathbf{c}_\mathrm{train}$) paired with synthetic measurements ($\mathbf{b}_\mathrm{train} = \mathbf{Ac}_\mathrm{train}$) are used to train the network. Phantoms for training have been obtained from previous reconstructions~\cite{Huang2019,Huang2020}, random Gaussian fields~\cite{Huang2018,Wei2020}, and large-eddy simulations~\cite{Wei2021}; random errors are usually added to $\mathbf{b}_\mathrm{train}$ to make the reconstruction procedure robust to noise. While this approach can yield accurate estimates of the QoI when the training set accurately represents the target physics, the application of DNN-based reconstruction to targets that exhibit unique flow structures is not reliable.\par

Deep learning has been used to post-process conventional tomographic estimates to improve the accuracy and resolution of reconstructed fields as well as to infer additional QoI. Notably, the group of Karniadakis reconstructed the temperature field induced by natural convection above a hot espresso cup via BOS tomography. The authors then fed their reconstructions to a PINN to both refine the temperature field as well as determine the velocity and pressure fields~\cite{Cai2021a}. Unfortunately, by starting from error-laden reconstructions, this technique fails to fully capitalize on the available information, as substantiated in section~\ref{sec:demo}. Apart from \cite{Cai2021a}, several groups developed CNNs to increase the resolution of flow fields in post-processing, e.g., \cite{Ferdian2020,Gao2021}, but these examples relied upon supervised training and CNNs are less effective than PINNs at incorporating flow physics into the estimation procedure~\cite{Cai2021d}.\par

Cai et al.~\cite{Cai2021b} took an initial step beyond PINN-based post-processing with their development of artificial intelligence velocimetry (AIV). The authors used a PINN to estimate 2D and 3D flow fields in a microfluidic channel from a series of images that were recorded with a single camera. The PINN outputted image intensities that were compared to experimental images in a data loss term. Additional losses were included to enforce a no-slip boundary condition along the channel walls as well as the Navier--Stokes equations throughout the domain. Unlike CFD-based analysis, AIV does not rely on inlet or outlet boundary conditions. Flow fields produced by AIV were comparable to estimates from other methods such as Deep-PIV~\cite{Cai2019b}, optical flow~\cite{Heitz2010}, and manual platelet tracking. However, the authors did not examine the influence of noise on the reliability of inferred fields. Moreover, the data loss term in AIV did not account for the LoS-integrated nature of the image data, which is a crucial consideration in flow field tomography and is central to our direct methodology.\par

\subsection{Direct Reconstruction with a PINN}
\label{sec:tomography:PINN algorithm}
Physics-informed neural nets employ a simple DNN architecture to solve both forward and inverse problems that are governed by one or more partial differential equations (PDEs) given a limited number of observations~\cite{Raissi2019}. PINNs map spatio-temporal coordinates that are fed into the input layer to the corresponding QoI at the output layer through a deep, fully-connected, feedforward network; an example of this architecture can be seen in Fig.~\ref{fig:PINN architecture}. The network is essentially a functional representation of the relationship between inputted spacetime coordinates and the unknown variables. In a typical deep learning scenario, the outputs of a DNN are compared to known values (training data), and the network's parameters (weights and biases) are tuned via backpropagation (BP) to minimize the residuals, which collectively form a ``data loss'' term. BP uses automatic differentiation (AD) to compute partial derivatives of the outputs with respect to the weights, biases, and inputs of a network~\cite{Baydin2018}. PINNs take advantage of this information to provide an additional loss term: partial derivatives of the outputs with respect to the inputs are plugged into the governing PDE(s) and the residuals are aggregated to form an overall ``physics loss.'' Simultaneously minimizing the data and physics loss terms via BP promotes outputs that both match the data from measurements or simulations and conform to known physics.\par

Raissi et al.~\cite{Raissi2020} utilized PINNs to estimate 2D and 3D fluid flow fields from sparse data. Here, we consider a 2D case in which $(x,y,t)$ coordinates are mapped to the flow fields of interest, $(c,u,v,p)$, where $c$ is the concentration of a passive scalar, $u$ and $v$ are the $x$- and $y$-direction components of velocity, and $p$ is pressure. In this context, fluid motion is governed by the incompressible Navier--Stokes equations as well as an advection--diffusion equation. These PDEs are written in non-dimensional form and rearranged to yield a set of ``physics residuals,''
\begin{subequations}
\begin{align}
    e_1 &= u_x + v_y \label{equ:physics residuals:mass}\\
    e_2 &= u_t + u\,u_x + v\,u_y + p_x - Re^{-1}\left(u_{xx} + u_{yy}\right) \label{equ:physics residuals:x momentum}\\
    e_3 &= v_t + u\,v_x + v\,v_y + p_y - Re^{-1}\left(v_{xx} + v_{yy}\right) \label{equ:physics residuals:y momentum}\\
    e_4 &= c_t + u\,c_x + v\,c_y - Pe^{-1}\left(c_{xx} + c_{yy}\right), \label{equ:physics residuals:scalar advection}
\end{align}
\label{equ:physics residuals}%
\end{subequations}
where $Re$ is the Reynolds number, $Pe$ is the P{\'e}clet number, and the subscripts $\cdot_x$, $\cdot_y$, and $\cdot_t$ indicate a partial derivative with respect to space or time. Residuals at each node are organized into a vector at each timestep. That is, $\mathbf{e}^{(k)} = \{e_{1,j}^{(k)}, e_{2,j}^{(k)}, e_{3,j}^{(k)}, e_{4,j}^{(k)}\}_{j=1}^n$ contains the values of \eqref{equ:physics residuals:mass}--\eqref{equ:physics residuals:scalar advection} at each node at the $k$th timestep, and a matrix of residuals is defined as follows:
\begin{equation}
    \mathbf{E} = [\mathbf{e}^{(1)}, \mathbf{e}^{(2)},\dots, \mathbf{e}^{(K)}],
    \label{equ:physics residual matrix}
\end{equation}
where $K$ is the total number of timesteps used for training, as in \eqref{equ:QoI matrix} and \eqref{equ:measurement matrix}. Finally, the physics loss term is simply
\begin{equation}
    \mathcal{L}_\mathrm{phys} = \left\lVert \mathbf{E} \right\rVert_\mathrm{F}^2,
    \label{equ:physics loss}
\end{equation}
where $\lVert\cdot\rVert_\mathrm{F}$ denotes the Frobenius norm.\par

In a previous effort, the data loss term was based on ``known'' values of some QoI at selected $(x,y,z,t)$ points in the flow field~\cite{Cai2021a}. For instance, the concentration field outputted by the PINN, $\mathbf{C}_\mathrm{out}$, could be compared to reconstructions computed using one of the conventional algorithms mentioned in section~\ref{sec:tomography:explicit algorithms}, denoted $\mathbf{C}_\mathrm{est}$:
\begin{equation}
    \mathcal{L}_\mathrm{est} = \left\lVert \mathbf{C}_\mathrm{out} - \mathbf{C}_\mathrm{est} \right\rVert_\mathrm{F}^2.
    \label{equ:reconstruction loss}
\end{equation}
However, $\mathbf{C}_\mathrm{est}$ is subject to reconstruction errors that ultimately corrupt the function learned by minimizing this loss. As an alternative, we propose to incorporate the measurement model from \eqref{equ:model discrete} into the loss function,\footnote{Similar methods have been used to reconstruct blood flow fields from magnetic resonance imaging data~\cite{VanHerten2020,Fathi2020} and wind fields from light detection and ranging data~\cite{Zhang2021}. Our study represents the first such use of PINNs for tomographic reconstruction.}
\begin{equation}
    \mathcal{L}_\mathrm{meas} = \left\lVert \mathbf{AC}_\mathrm{out} - \mathbf{B}\right \rVert_\mathrm{F}^2,
    \label{equ:measurement loss}
\end{equation}
which could be done for most of the modalities listed in the introduction.\footnote{Complications arise in tomographic PIV because the projections are proportional to LoS integrals over discrete particle fields as opposed to integrals of the QoI.} In other words, a PINN trained with \eqref{equ:measurement loss} will learn a function that satisfies the governing physics and the \textit{projection data}, as opposed to error-prone reconstructions. Finally, the loss terms are weighted and combined to form a total loss,
\begin{subequations}
\begin{align}
    \mathcal{L}_\mathrm{total} &= \gamma\mathcal{L}_\mathrm{est} + \mathcal{L}_\mathrm{phys} \quad\text{or} \label{equ:total loss:reconstruction}\\
    \mathcal{L}_\mathrm{total} &= \gamma\mathcal{L}_\mathrm{meas} + \mathcal{L}_\mathrm{phys}, \label{equ:total loss:measurement}
\end{align}
\label{equ:total loss}%
\end{subequations}
where the parameter $\gamma$ determines the relative influence of physics and data losses. Equations \eqref{equ:total loss:reconstruction} and \eqref{equ:total loss:measurement} are used for post-processing (per \cite{Cai2021a}) and direct reconstruction, respectively.\par

\subsection{Reconstruction Errors}
\label{sec:tomography:errors}
We quantify errors in flow field reconstructions in terms of a normalized Euclidean distance,
\begin{equation}
    \varepsilon_\mathrm{x} = \frac{\left\lVert \mathbf{x}_\mathrm{exact} - \mathbf{x} \right\rVert_2^2}{\left\lVert \mathbf{x}_\mathrm{exact} \right\rVert_2^2},
    \label{equ:error}
\end{equation}
where $\lVert\cdot\rVert_2$ is a Euclidean norm; $\mathbf{x}_\mathrm{exact}$ is a vector of concentration ($\mathbf{c}$), velocity ($\mathbf{u}$ or $\mathbf{v}$), or pressure ($\mathbf{p}$) data from a known ground truth field; and $\mathbf{x}$ is the corresponding estimate.\par

\section{2D Demonstration}
\label{sec:demo}
\begin{figure}[t]
    \centering
    \includegraphics[width=.475\textwidth]{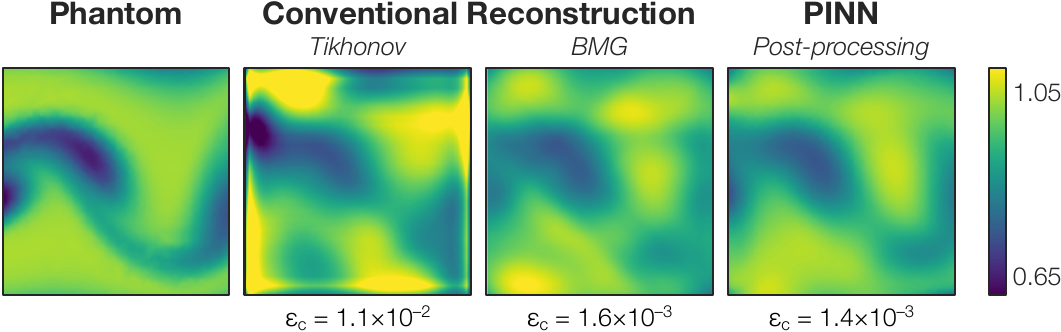}
    \caption{Reconstructions of a 2D flow field from noise-free data. Conventional (i.e., non-PINN) reconstructions were obtained by Tikhonov regularization and the Bayesian algorithm described in \cite{Grauer2019}. The latter reconstructions were post-processed with a PINN using \eqref{equ:total loss:reconstruction}, which provided a moderate reduction in concentration field errors as well as velocity and pressure field estimates.}
    \label{fig:conventional reconstructions}
\end{figure}

We demonstrate our approach to physics-informed flow field tomography with a 2D example that resembles absorption-based modalities~\cite{Cai2017}. The target flow contains a passive scalar that is transported over a cylindrical bluff body, resulting in an unsteady wake of vortices travelling in the streamwise direction. Concentration, velocity, and pressure data for a $Re = 100$ and $Pe = 100$ flow was obtained from the direct numerical simulation (DNS) of Raissi et al.~\cite{Raissi2020}. Their computational domain consisted of 30,189 nodes distributed throughout a $20\times40$ (dimensionless) area, and they provided data for 201 timesteps. We selected a $2\times2$ interrogation region behind the cylinder having 6561 nodes. A snapshot of the DNS concentration field in this region can be seen in Fig.~\ref{fig:conventional reconstructions}.

Readers should note that this work is limited by the use of a single flow scenario. However, PINNs have learned a wide variety of flow fields, discussed in depth by Cai et al.~\cite{Cai2021d}, including turbulent channel flow~\cite{Jin2021}, pulsatile blood flow through arteries~\cite{Kissas2020}, forced convective flow within a power electronics enclosure~\cite{Cai2021c}, 2D flows with bow and oblique shock waves~\cite{Mao2020}, and flow induced by natural convection~\cite{Cai2021a}. Successful PINN representation of these flows suggest that, given a proper measurement model and sufficient projection data, PINN-based flow field tomography could be deployed in comparable scenarios. The following tests serve to explicate the training properties of PINNs with a tomographic measurement model.\par

\subsection{Measurements and Grid}
\label{sec:demo:grid}
\begin{figure}[t]
    \centering
    \includegraphics[width=.495\textwidth]{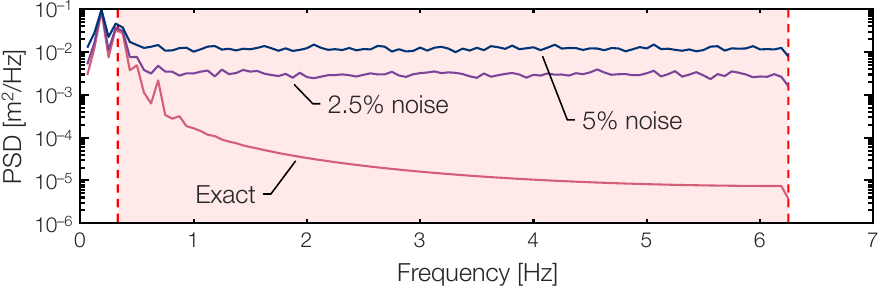}
    \caption{Average PSD of exact projections from the 100-beam data set as well as projections corrupted with 2.5\% and 5\% noise. Note that the high-frequency signal is almost entirely occluded by noise, which has a magnitude on par with the low-frequency signal in the case of 5\% errors. This is due to the method for adding noise, i.e., 2.5\% and 5\% are relative to the largest single measurement from any beam across the whole data set.}
    \label{fig:noise}
\end{figure}

Synthetic projection data was generated using path integrals of the concentration field along ``beams.'' Measurements were calculated by high-order quadrature to introduce discrepancies between exact measurements of the quasi-continuous flow field and discrete projections of the DNS data, $\mathbf{Ac}_\mathrm{exact}$, to avoid the ``inverse crime''~\cite{Wirgin2004}. All of the beam arrays featured four viewing angles with a 45$^\circ$ separation between adjacent views. Beams in each view were evenly spaced across the domain, and we tested arrangements with 20, 40, 100, and 160 beams; the 20- and 100-beam arrays are shown in Fig.~\ref{fig:array}. In addition to idealized noise-free data, we reconstructed measurements corrupted by independent and identically distributed (IID) Gaussian errors having a standard deviation equal to 2.5\% or 5\% of $\mathrm{max}(\mathbf{B}_\mathrm{exact})$, resulting in individual projection errors up to 25\% in the 2.5\% case and 56\% in the 5\% case. We colloquially refer to these errors as noise throughout the text. Figure~\ref{fig:noise} depicts the average power spectral density of the noise-free and noise-laden 100-beam data sets. The magnitude of noise is evident in this figure: the high-frequency signal is dominated by high-frequency noise, which is commensurate with the low-frequency signal in the 5\% case.\par

\begin{figure*}[b]
    \centering\hfill
    \subfloat[]{
        \includegraphics[width=.475\textwidth]{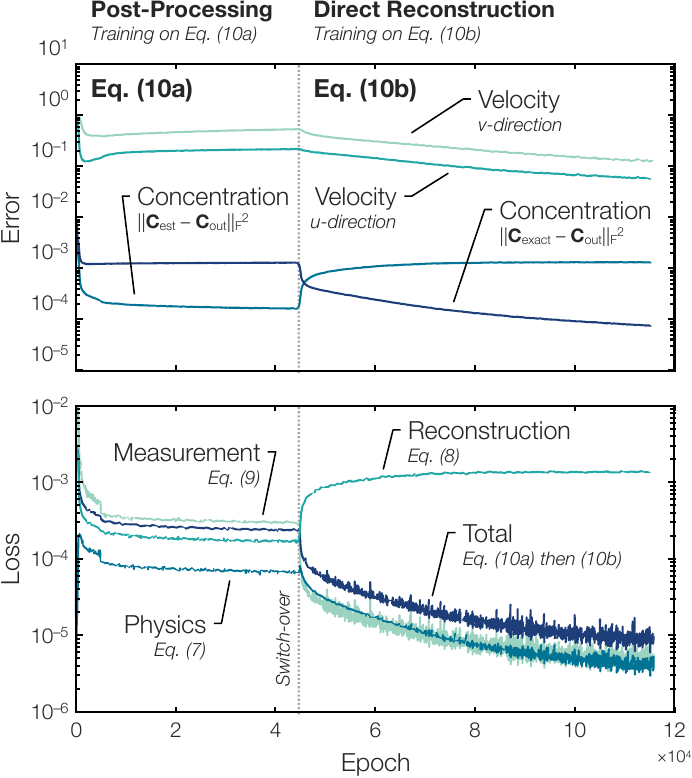}
        \label{fig:training:switch-over}}\quad
    \subfloat[]{
        \includegraphics[width=.475\textwidth]{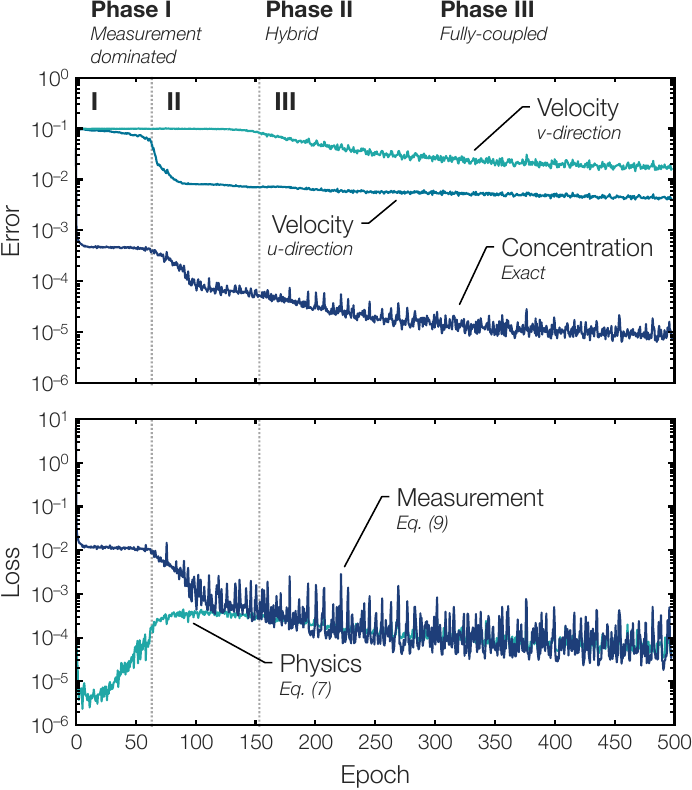}
        \label{fig:training:regimes}}\hfill\null
    \caption{Reconstruction errors and losses throughout training: (a) switch-over from post-processing to direct reconstruction and (b) phases of direct reconstruction; both tests feature noise-free data from 100 beams. In (a), the physics loss plateaus prior to the switch-over due to non-physical reconstruction errors in $\mathbf{C}_\mathrm{est}$. After switching to direct reconstruction, the PINN is able to considerably reduce the measurement and physics losses, and errors in all of the fields are minimized. In (b), three phases of direct reconstruction can be seen: (I) the concentration field is adjusted to minimize measurement errors while physics losses increase; (II) the concentration field improves while physics losses level-off, during which time the concentration and streamwise velocity fields improve; and (III) the measurement and physics losses are simultaneously minimized and the accuracy of all fields increases.}
    \label{fig:training}
\end{figure*}

Flow field variables were represented using the finite element method with a triangle element mesh, per \cite{Grauer2019}. We tested a low-dimensional grid with 1250 nodes, pictured in Fig.~\ref{fig:array}, and a high-dimensional grid with 3600 nodes. As such, the rank of $\mathbf{A}$ was considerably lower than the number of unknowns in each case, resulting in a limited-data imaging scenario.\par

\subsection{PINN Architecture and Training}
\label{sec:demo:training}
Physics-informed neural nets used in this work were implemented in both TensorFlow~\cite{Abadi2016} and PyTorch~\cite{Paszke2019}. The networks discussed in this section comprised ten hidden layers that contained 50 neurons per output variable. We used swish activation functions because they have been shown to improve the stability of the gradients needed to calculate \eqref{equ:physics residuals} compared to hyperbolic tangent and rectified linear activation functions~\cite{Raissi2020}. Weights were randomly initialized with a standard normal distribution and biases were set to zero at the start. Training was performed by minimizing \eqref{equ:total loss:reconstruction} or \eqref{equ:total loss:measurement} with the Adam optimizer~\cite{Kingma2014} at a learning rate of $1\times10^{-3}$ for the first 1000 epochs and $1\times10^{-4}$ thereafter, where an epoch indicates one pass through the full data set. The PINNs were trained until the total loss reached a plateau, defined as a 1000-epoch stretch over which the 500-epoch running average of $\mathcal{L}_\mathrm{total}$ decreased by less than 1\%. On average, reconstructions computed on an NVIDIA Tesla P100 graphis processing unit took 95 hours to satisfy this criterion.\par

\subsection{Post-Processing vs. Direct Reconstruction}
\label{sec:demo:comparison}
We first set out to compare the performance of a PINN trained on \eqref{equ:total loss:reconstruction}, i.e., \textit{post-processing} of a conventional reconstruction, to a PINN trained on \eqref{equ:total loss:measurement}, i.e., \textit{direct reconstruction} of the projection data.\par

\begin{figure*}[t]
    \centering
    \includegraphics[width=0.95\textwidth]{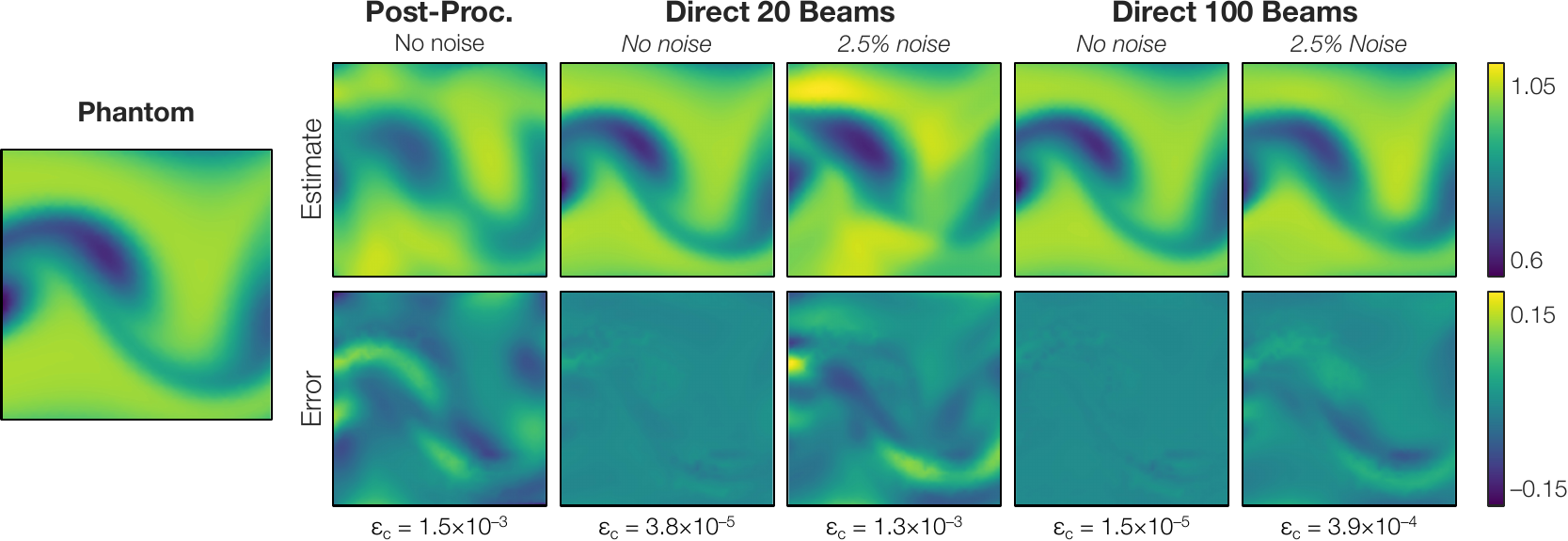}
    \caption{Comparison of concentration fields estimated from projection data using a PINN. A ground truth field is shown on the left; reconstructions and error fields are plotted to the right. The first column of estimates and errors shows the results of post-processing conventional reconstructions of noise-free projections from 100 beams; the remaining columns depict direct reconstructions. Directly reconstructing noisy data from 20 beams produced quantitatively and qualitatively better estimates of the concentration field than post-processing reconstructions obtained with more, cleaner data.}
    \label{fig:error panel}
\end{figure*}

Post-processing requires a suitable estimate of the scalar field to specify the data loss term ($\mathbf{C}_\mathrm{est}$ in \eqref{equ:reconstruction loss}), which can be obtained using one of the algorithms listed in section~\ref{sec:tomography:explicit algorithms}. To start, we reconstructed the scalar field with noise-free data from 100 beams using Tikhonov regularization, but we found that these estimates exhibited large errors and barely resembled the phantoms. One instance from the reconstructed set is shown to the right of the phantom in Fig.~\ref{fig:conventional reconstructions}. Next, we tried the Bayesian algorithm described by Grauer et al.~\cite{Grauer2019}, in which the flow is modeled as a multivariate Gaussian process with an assumed covariance structure. We adopted the exponential decay matrix from \cite{Grauer2019} and conducted a parametric study to optimize the variance and length scale parameters in the prior, i.e., to ensure best case scalar field reconstructions for post-processing. These estimates are hereafter referred to as Bayesian multivariate Gaussian (BMG) reconstructions. Figure~\ref{fig:conventional reconstructions} depicts a BMG reconstruction alongside the corresponding Tikhonov estimate. The BMG concentration field is visibly superior to the Tikhonov one, although both reconstructions contain sizable artifacts. We then post-processed the BMG fields with a PINN using \eqref{equ:total loss:reconstruction} by setting $\mathbf{C}_\mathrm{est}$ to be the Bayesian estimates; the final panel of Fig.~\ref{fig:conventional reconstructions} depicts a resultant concentration field from this procedure. Average concentration field errors were $1.1\times10^{-2}$ for the Tikhonov estimates, $1.6\times10^{-3}$ for the BMG reconstructions, and $1.4\times10^{-3}$ after post-processing the latter fields with a PINN. In addition to mitigating reconstruction errors in $\mathbf{C}_\mathrm{est}$, post-processing produced estimates of the velocity and pressure fields. However, several large-scale artifacts in the conventional estimates were left untouched by the PINN since those structures were treated as known by the data loss function in \eqref{equ:total loss:reconstruction}. As a result, the supplemental fields were also prone to significant inaccuracies.\par

\begin{figure}[t]
    \centering
    \includegraphics[width=.45\textwidth]{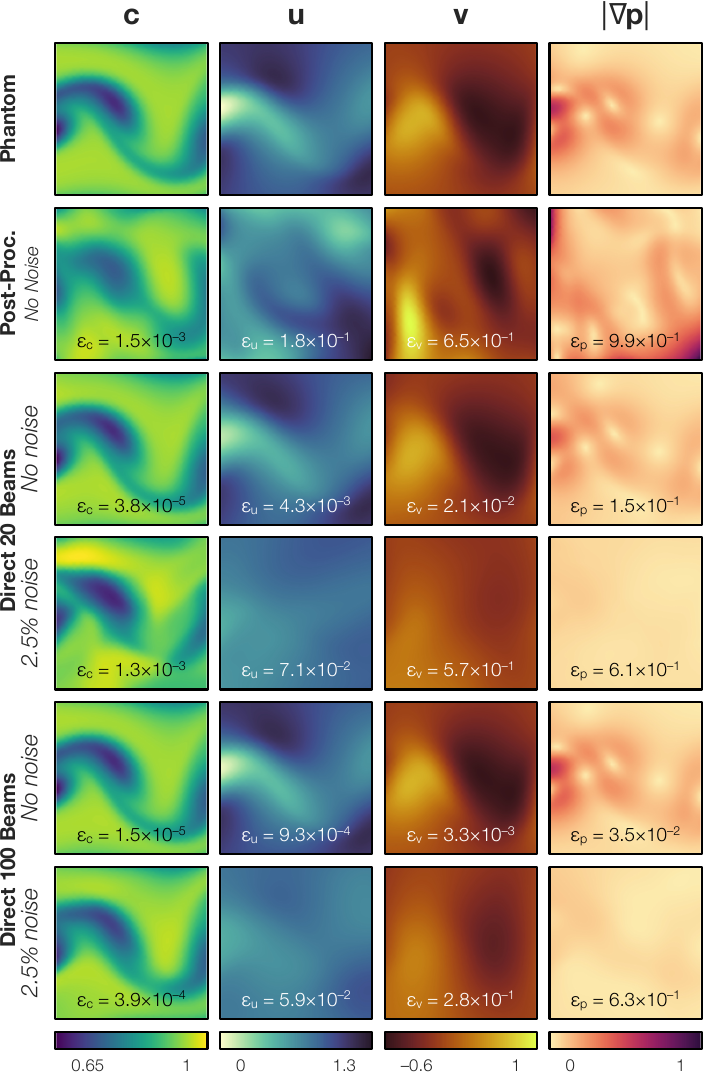}
    \caption{Concentration, velocity, and pressure gradient fields from a 2D flow: (row one) phantom distributions from a DNS, (row two) post-processing of BMG reconstructions of noise-free data from 100 beams, (row three) direct reconstruction of noise-free data from 20 beams, (row four) direct reconstruction of noisy (2.5\%) data from 20 beams, (row five) direct reconstruction of noise-free data from 100 beams, (row six) direct reconstruction of noisy (2.5\%) data from 100 beams. Errors in $\mathbf{C}_\mathrm{est}$ produced by the conventional (BMG) algorithm corrupt subsequent post-processing (row two); direct reconstruction yielded noticeably refined fields with less measurement data, even in the presence of noise.}
    \label{fig:CUVP panel}
\end{figure}

The difference between post-processing and direct reconstruction is effectively illustrated through the use of a switch-over from \eqref{equ:total loss:reconstruction} to \eqref{equ:total loss:measurement}, the latter of which utilizes the tomographic measurement model to compare the PINN's output to projection data instead of error-laden reconstructions. To test this hypothesis, we set up a PINN to minimize \eqref{equ:total loss:reconstruction} for 45,000 epochs, at which point the loss function was converted to \eqref{equ:total loss:measurement}. Figure~\ref{fig:training:switch-over} depicts average concentration and velocity field errors as well as losses throughout training. All fields quickly improved from their initial random state, stabilizing after approximately 10,000 epochs, at which point the PINN's output nearly matched the conventional Bayesian reconstructions. This can be seen in the comparison between $\mathbf{C}_\mathrm{est}$ (i.e., the BMG reconstructions) and $\mathbf{C}_\mathrm{out}$ in Fig.~\ref{fig:training:switch-over}. A striking change can be seen at the switch-over: the outputted concentration field departs from $\mathbf{C}_\mathrm{est}$ and rapidly approaches $\mathbf{C}_\mathrm{exact}$. This suggests that errors in $\mathbf{C}_\mathrm{est}$ are incompatible with the flow field physics and produce a high physics residual that cannot be overcome due to the trade-offs between satisfying \eqref{equ:physics loss} and \eqref{equ:reconstruction loss}. During the direct reconstruction stage, physics losses fell far below their plateau at the end of post-processing, errors in all fields were considerably reduced after the switch-over. Average errors in the post-processed fields were $\varepsilon_\mathrm{c} = 1.4\times10^{-3}$, $\varepsilon_\mathrm{u} = 1.8\times10^{-1}$, $\varepsilon_\mathrm{v} = 6.8\times10^{-1}$, and $\varepsilon_\mathrm{p} = 7.3\times10^{-1}$. Errors in the directly reconstructed fields were $\varepsilon_\mathrm{c} = 1.9\times10^{-5}$, $\varepsilon_\mathrm{u} = 6.3\times10^{-3}$, $\varepsilon_\mathrm{v} = 2.7\times10^{-2}$, and $\varepsilon_\mathrm{p} = 1.1\times10^{-1}$. This amounts to substantial decreases in error of 99\%, 97\%, 96\%, and 85\%, respectively!\par

In order to verify that direct reconstruction could be conducted without a conventional precursor (that is to say, without training a PINN on \eqref{equ:total loss:reconstruction} to predict $\mathbf{C}_\mathrm{est}$ before switching to measurement losses), we trained another PINN on \eqref{equ:total loss:measurement} from the outset. Figure~\ref{fig:training:regimes} depicts the errors and losses recorded at the start of this test. We observed three distinct regimes of training behavior. First was a \textit{measurement dominated regime} (phase I), during which the concentration field was quickly altered to satisfy the projection data, i.e., minimize $\mathcal{L}_\mathrm{meas}$. As with conventional tomography, minimizing the measurement loss per se is insufficient for reconstruction since there are infinitely many concentration fields that satisfy the projection data, most of which are non-physical. Moreover, there were no meaningful changes to the velocity fields nor to the pressure field during the first regime. In fact, the physics losses, which began at a very low value, increased throughout the first regime since the concentration field developed significant structure while the other fields remained effectively random. After several dozen epochs, the physics losses leveled off (but did not appreciably improve) and the measurement losses started to decrease once again. The peak of $\mathcal{L}_\mathrm{phys}$ marked the beginning of a \textit{hybrid regime} (phase II) during which errors in the streamwise velocity field started to diminish. Measurement losses approached the physics losses after around 200 epochs of training, indicating the onset of a \textit{fully-coupled regime} (phase III). This also coincided with maximum continuity and momentum residuals, i.e., \eqref{equ:physics residuals:mass}--\eqref{equ:physics residuals:y momentum}, since there was no meaningful progress in the transverse velocity field to accompany new structure in $u$ from phase II. All flow fields were simultaneously improved and all losses simultaneously minimized during the third and final phase of training, which continued until convergence.\par

We note two important aspects of the results in Fig.~\ref{fig:training}. First, the flow fields produced at the end of switch-over test were virtually identical to those produced by the PINN trained solely on \eqref{equ:total loss:measurement}, which suggests that curriculum learning is not necessary for direct reconstruction. Second, the training regimes in Fig.~\ref{fig:training:regimes} were observed in every noise-free case that we tested. The effects of noise are significant, however, and we discuss them in detail in the next section, along with our procedure for reconstructing noisy data with a C-PINN. \par

Figure~\ref{fig:error panel} depicts a snapshot of the DNS-based ground truth concentration field alongside a series of estimates obtained by post-processing and direct reconstruction. Post-processing was conducted using BMG reconstructions of noise-free projections from 100 beams, as before. Also shown are direct reconstructions of projections from 20 beams and 100 beams, both with noise and without. Errors fields are plotted below each reconstruction and the Euclidean error is listed below that. Figure~\ref{fig:CUVP panel} shows the corresponding velocity and pressure gradient fields next to the same set of concentration fields.\footnote{Throughout this paper, we plot $|\nabla p|$ instead of $p$ since the physics residuals feature $p_x$ and $p_y$ and a reference pressure is required to infer $p$. Errors in $|\nabla p|$ are denoted $\varepsilon_\mathrm{p}$.} In our switch-over test, training the PINN on \eqref{equ:total loss:reconstruction} produced the ``post-proc.'' fields and subsequent training on \eqref{equ:total loss:measurement} produced the ``direct 100 beams, no noise'' fields. The quantitative reduction in error plotted in Fig.~\ref{fig:training:switch-over} corresponds to the qualitative improvement in estimates of $c$, $u$, $v$, and $|\nabla p|$ shown in Figs.~\ref{fig:error panel} and \ref{fig:CUVP panel}. Direct reconstruction was far more accurate than post-processing. Indeed, there was a near-perfect visual correspondence of all fields directly reconstructed from 20 noise-free projections to the ground truth flow fields as well as a reduction in error of two orders of magnitude relative to the post-processed results. Moreover, direct reconstructions of $v$ and $|\nabla p|$ from 20 noisy projections were comparable to their post-processed counterparts, despite the availability of five times more data with zero measurement errors in the latter case. In all our tests, the concentration fields derived from post-processing were noticeably corrupted by the reconstruction errors in $\mathbf{C}_\mathrm{est}$, and the velocity and pressure gradient fields bore little resemblance to the phantoms. This further confirms our supposition that reconstruction errors in $\mathbf{C}_\mathrm{est}$ are incompatible with the physics residuals returned by \eqref{equ:physics residuals}.\par

\subsection{Measurement Error and Semi-Convergence}
\label{sec:demo:semi-convergence}
Idealized projections, $\mathbf{b}_\mathrm{exact} = \mathbf{Ac}_\mathrm{exact}$, are never available in practice, and the inevitable discrepancies between $\mathbf{b}_\mathrm{exact}$ and real data can have appreciable adverse effects on a reconstruction. Such discrepancies arise due to model errors, e.g., stemming from the approximation in \eqref{equ:model continuous}, beam misalignment or calibration errors, laser intensity fluctuations, read noise, thermal noise, and so on. Hence, it is crucial to assess the stability of reconstructions in the presence of measurement errors.\par

\begin{figure}[t]
    \centering
    \includegraphics[width=.475\textwidth]{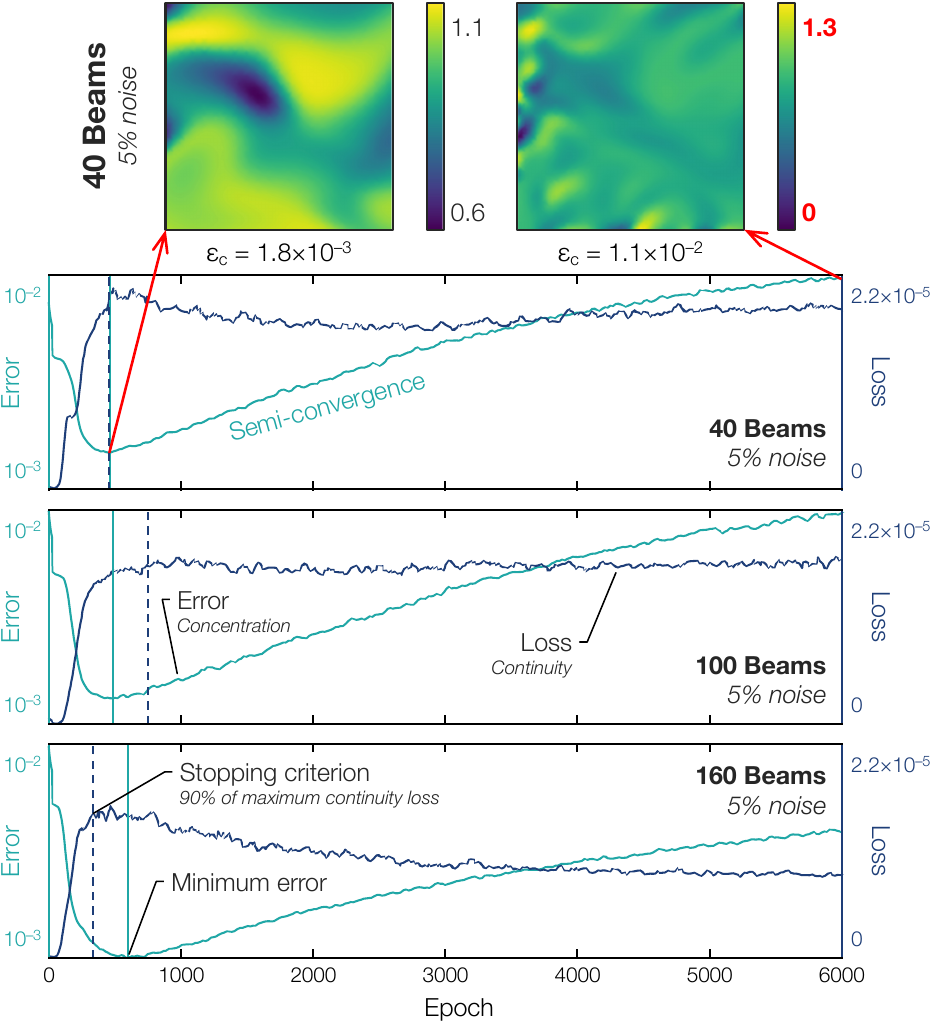}
    \caption{Semi-convergence was observed in the presence of measurement noise regardless of the number of projections. A stopping criterion was determined using the reconstruction phases observed in Fig.~\ref{fig:training:regimes}. Network parameters were extracted from the point at which a five-epoch running average of the continuity residual reached 95\% of its maximum value, which roughly corresponds to the transition from phase I to II. This procedure yielded a good approximation to the true concentration error minimum. It should be noted that training was conducted at a fixed learning rate of $1\times10^{-3}$ to generate these figures (i.e., instead of switching to $1\times10^{-4}$ after 1000 epochs), although this did not materially affect our results.}
    \label{fig:semi-convergence}
\end{figure}

When applied to noisy (or otherwise error-laden) data, iterative reconstruction algorithms such as the additive ART, MART, and SIRTs usually approach the exact solution for a few steps before changing course towards a relatively poor estimate~\cite{Elfving2014}. This behavior is called semi-convergence and has been the subject of numerous theoretical and numerical analyses, e.g., \cite{Hansen2018,Elfving2014,Natterer2001,Censor2008,Byrne2019}. We observed a similar phenomenon during the direct reconstruction of noisy projections with a PINN. Figure~\ref{fig:semi-convergence} contains concentration field error and continuity loss traces from three tests, using projections from arrays with 40, 100, and 160 beams; the data in these tests were corrupted with 5\% IID Gaussian errors, per section~\ref{sec:demo:grid}. In each case, there was a progressive decrease in $\varepsilon_\mathrm{c}$ for approximately 500 epochs after which the concentration field errors began to increase. The qualitative effect of this trend can be seen in the sample 40-beam reconstructions located above the trace plots. The left-most concentration field was outputted near the point of semi-convergence, i.e., the minimum value of $\varepsilon_\mathrm{c}$, while the snapshot on the right was outputted at 6000 epochs, well after semi-convergence. The former reconstruction is similar to the phantom (cf. the exact concentration field shown in Figs.~\ref{fig:conventional reconstructions}--\ref{fig:CUVP panel}) whereas the latter estimate contains significant artifacts (note the larger data range).\par

Conventional iterative reconstruction algorithms exhibit semi-convergence because early iterations add components to $\mathbf{c}$ that have a low spatial frequency while later contributions feature high-frequency components that are more easily influenced by errors in $\mathbf{b}$ (see the PSDs in Fig.~\ref{fig:noise}). This tendency can be deduced through a singular value decomposition of $\mathbf{A}$, which is left-multiplied by the corresponding preconditioner matrix in the case of a SIRT algorithm~\cite{Elfving2014}. It so happens that most flow fields of interest have significant low-frequency content, resulting in semi-convergence, although this is not mathematically necessary (meaning that one can devise a phantom which does not lead to semi-convergence). Crucially, the measurement residual, $\lVert\mathbf{Ac}^{(k)} - \mathbf{b}\rVert_2^2$, continues to monotonically decrease with further iterations in spite of semi-convergence. In other words, the fully-converged, error-prone reconstruction accurately reflects the information implicit in the objective function, subject to noisy data. In principle, this can be overcome through the use of an objective function that contains more comprehensive and/or accurate information, which is the aim of regularization and motivates our use of $\mathcal{L}_\mathrm{phys}$.\par

A similar semi-convergence mechanism is at work in tomographic reconstruction with a PINN. Resembling the iterative linear solvers mentioned above, DNNs trained by gradient descent have been shown to learn low-frequency components of the outputs early on, followed by components of increasing frequency as training continues~\cite{Ronen2019,Basri2020}. In particular, an analysis of PINNs revealed a spectral bias in training as well as differential rates of convergence among various components of the loss term~\cite{Wang2020b}. As a result, early truncation of training essentially amounts to frequency-based regularization of the network's outputs. This effect can be seen in Fig.~\ref{fig:semi-convergence}, wherein the erratic, inaccurate estimate on the right side yielded a lower overall loss than the more reasonable, semi-converged estimate on the left side because the fully-converged PINN was overfitted to high-frequency components of the data that were dominated by noise. It is interesting to note that this behavior also appeared in our switch-over test: velocity field errors during post-processing can be seen to exhibit semi-convergence in Fig.~\ref{fig:training:switch-over}, with errors in $\mathbf{u}$ and $\mathbf{v}$ subsequently increasing as the PINN matched its output to $\mathbf{C}_\mathrm{est}$.\par

\begin{figure}[t]
    \centering
    \includegraphics[width=.45\textwidth]{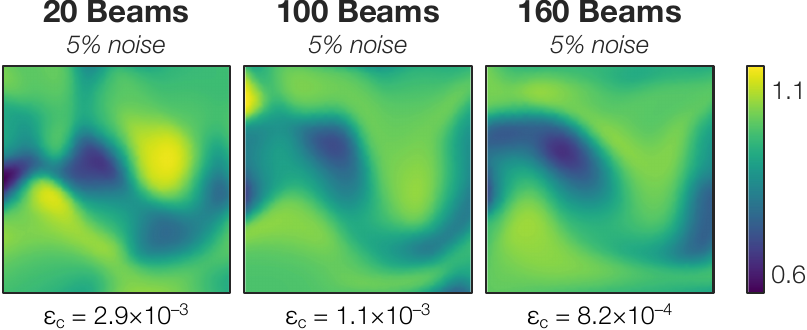}
    \caption{While reconstructions are adversely affected by measurement noise, accurate concentration field estimates can be obtained using the maximum continuity residual stopping criterion given a sufficient number of projections.}
    \label{fig:noise test}
\end{figure}

One obvious solution to the problem of semi-convergence, akin to classical regularization, is to place greater emphasis on the physics residuals by decreasing $\gamma$. However, this did not produce the intended result, as shown in sections~\ref{sec:demo:gamma}, \ref{sec:demo:performance}, and \ref{sec:Bayesian}. Therefore, given noisy data, it is desirable to halt training prior to convergence, i.e., to use iterative regularization. It is not desirable, however, to select the halting point heuristically (although doing so is common practice in conventional flow field tomography). Fortunately, the training regimes introduced above can be used to time the iterative regularization of direct reconstructions. Recall that the maximum continuity and momentum losses occurred at the transition from phase I to II of training on noise-free data, where phases I and II were relatively quick compared to the final, fully-coupled phase. While this trend is readily apparent by visual inspection, as in the continuity traces in Fig.~\ref{fig:semi-convergence}, the boundary between phases is somewhat vague so the details of any precise metric will be arbitrary. We extracted the network parameters at the point where a five-epoch average of the continuity loss reached 95\% of its maximum value. This criterion neatly coincided with the physics loss plateau in all of our tests.\par

The location of semi-convergence and our stopping point are indicated in Fig.~\ref{fig:semi-convergence} by a solid turquoise line and dashed blue line, respectively. In each case, the maximum continuity loss was nearby the point of semi-convergence, sometimes occurring beforehand and other times afterwards. We conducted extensive tests using noisy data, outlined in section~\ref{sec:demo:performance}, throughout which the value of $\varepsilon_\mathrm{c}$ at our stopping point remained within $8.89\times10^{-4}$ of the true minimum, which is not known in practice; the average difference between $\varepsilon_\mathrm{c}$ at our stopping point and at semi-convergence was $2.72\times10^{-4}$. Several sample concentration fields obtained using our criterion are shown in Fig.~\ref{fig:noise test}, which depicts reconstructions of 20-, 100-, and 160-projection data sets contaminated by 5\% noise (the 40-beam estimate was shown earlier in Fig.~\ref{fig:semi-convergence}).\par

It is important to note that measurement errors have a greater impact on fields inferred solely from $\mathcal{L}_\mathrm{phys}$ ($u$, $v$, and $p$ in this case) than on fields that are also included in the data loss ($c$). This effect can be seen in Fig.~\ref{fig:CUVP panel}, where increases in $\varepsilon_\mathrm{u}$, $\varepsilon_\mathrm{v}$, and $\varepsilon_\mathrm{p}$ from a noise-free case to its noisy counterpart are considerably larger than the corresponding increase in $\varepsilon_\mathrm{c}$. Incorporating additional measurements that are sensitive to the velocity and pressure fields can dramatically improve all of the reconstructed fields.\par

\subsection{Optimizing $\gamma$}
\label{sec:demo:gamma}
\begin{figure}[t]
    \centering
    \includegraphics[width=.475\textwidth]{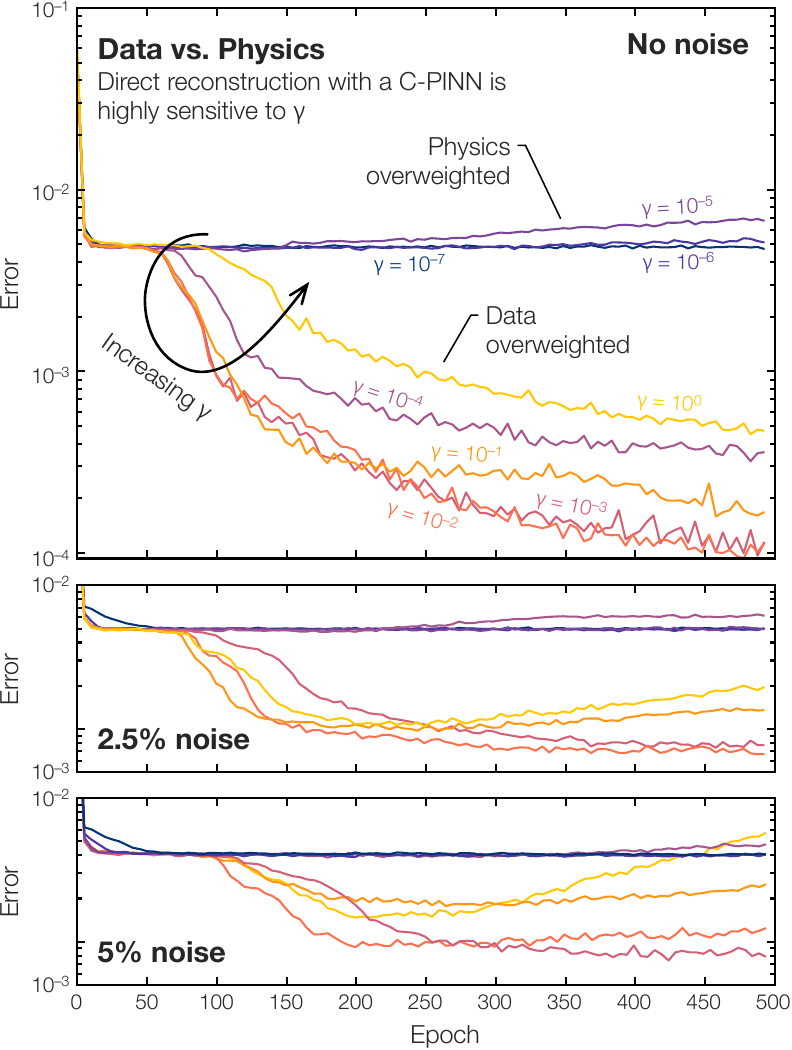}
    \caption{Concentration field errors during training for various weightings of the measurement and physics losses. Direct reconstruction with a PINN using conventional training is highly sensitive to the weighting parameter, $\gamma$. Over-emphasizing either the measurement or physics residuals can produce large errors in $c$, $u$, $v$, and $p$ due to intense local minima in $\mathcal{L}_\mathrm{total}$.}
    \label{fig:gamma test}
\end{figure}

The relative contribution of measurement and physics residuals to $\mathcal{L}_\mathrm{total}$ is an important consideration in direct reconstruction. We tested eight decibels of the weighting parameter, $\gamma$, ranging from $10^{-7}$ to $10^{0}$, in order to determine the optimal trade-off between $\mathcal{L}_\mathrm{meas}$ and $\mathcal{L}_\mathrm{phys}$. Testing was conducted on 100 clean projections as well as measurements perturbed by 2.5\% and 5\% noise. Average concentration field errors recorded in the first 500 epochs of training are shown in Fig.~\ref{fig:gamma test}. While exact $c$, $u$, $v$, and $p$ fields should always represent a significant minima of $\mathcal{L}_\mathrm{total}$ irrespective of the weighting parameter, especially in a noise-free test, values of $\gamma$ below $10^{-3}$ and above $10^{-1}$ failed to generate usable reconstructions. A low weight on the data loss tended to yield trivial solutions (for instance, $\mathcal{L}_\mathrm{phys}$ is minimized by zeros), and data-heavy weighting schemes produced reconstructions akin to those from a traditional iterative algorithm, like the ART, sans regularization. Weights of $\gamma = 10^{-3}$ and $10^{-2}$ produced acceptable reconstructions regardless of the level of noise. Slightly better performance was realized when $\gamma$ was set to $10^{-3}$ so we used this value in our tests unless otherwise mentioned.\par

Reconstructing noise-free data with $\gamma = 10^{-3}$ led to highly-accurate estimates of all fields, even in the 20-beam case, as can be seen in Fig.~\ref{fig:CUVP panel}. However, error-laden data resulted in semi-convergence, as described above, limiting the veracity of reconstructions. Accuracy of the velocity and pressure gradient fields was acutely limited because these fields were only learned by minimizing $\mathcal{L}_\mathrm{phys}$, and this process had to be prematurely truncated in order to regularize the concentration field. It stands to reason that, given an appropriate estimate of $c$, increasing the influence of $\mathcal{L}_\mathrm{phys}$ relative to $\mathcal{L}_\mathrm{meas}$ could counteract the effects of noise; e.g., $\gamma$ could be decreased near the point of semi-convergence. Changing the loss function in this way is known as adaptive training.\par

Several methods of adaptive training have been devised for PINNs, which could be applied in the case of direct reconstruction. For instance, Wang et al.~\cite{Wang2020a} developed a learning rate annealing procedure that automatically weights individual loss terms as a function of their gradient statistics. Several new parameters were introduced to control this process so Jin et al.~\cite{Jin2021} proposed a modification of Wang's method that required fewer user-specified inputs. We explored the potential of dynamic weighting by implementing step-wise reductions of $\gamma$ as well as an exponential decay of $\gamma$; adjustments to $\gamma$ were made at an idealized, predetermined point shortly after semi-convergence. Three step sizes were assessed, $\gamma_2/\gamma_1 = 10^{-1}$, $10^{-2}$, and $10^{-3}$; the exponential decay was conducted at a rate constant of unity; and both the step and decay tests were done with 2.5\% as well as 5\% noise. In several of these tests, there was a brief decrease in $\varepsilon_\mathrm{c}$, but the minimum value never fell below the semi-converged value. Moreover, most cases did not yield an appreciable decrease in error, and all the revised weighting schemes exhibited a secondary instance of semi-convergence, meaning that iterative regularization would be still be required when using an adaptive weighting scheme. Errors in the velocity and pressure gradient fields did not improve after re-weighting. These results do not support the use of adaptive weighting in direct reconstruction with a PINN. Moreover, the results in section~\ref{sec:Bayesian} suggest that the sensitivity to $\gamma$ observed in Fig.~\ref{fig:semi-convergence} is an artifact of the topology of $\mathcal{L}_\mathrm{total}$. Therefore, while it is possible to identify an appropriate value of $\gamma$ for conventional training, the statistical reconstruction framework introduced in section~\ref{sec:Bayesian} may be more suitable in the presence of measurement noise.\par

\subsection{Parametric Performance Tests}
\label{sec:demo:performance}
We conducted extensive testing to assess the effects of the grid resolution, beam count, and magnitude of measurement errors on the accuracy of direct reconstructions. Like in previous sections, post-processing was performed on BMG reconstructions of noise-free data from 100 beams to provide a baseline for comparison. Measurements from 20, 40, 100, and 160 beams were directly reconstructed, without noise to start and subsequently with 2.5\% and 5\% Gaussian errors. Moreover, every beam configuration and level of noise was tested in conjunction with our 1250-node ``coarse'' grid and the 3600-node ``fine'' grid. Average values of $\varepsilon_\mathrm{c}$, $\varepsilon_\mathrm{u}$, $\varepsilon_\mathrm{v}$, and $\varepsilon_\mathrm{p}$ from these tests are listed in table~\ref{tab:errors} along with concentration field errors of the BMG estimates.\par

\begin{table*}[t]
    \caption{Average reconstruction errors from parametric tests of direct reconstruction with a PINN}
    \centering
    \begin{tabular}{c c r r r r r r r r r}
     \hline\hline 
     \\[-2ex]
     \multirow{2}{*}{Noise} & \multirow{ 2}{*}{Beams} & \multicolumn{3}{c}{$\varepsilon_\mathrm{c} \times 10^{-5}$} & \multicolumn{2}{c}{$\varepsilon_\mathrm{u} \times 10^{-3}$} & \multicolumn{2}{c}{$\varepsilon_\mathrm{v} \times 10^{-2}$} & \multicolumn{2}{c}{$\varepsilon_\mathrm{p} \times 10^{-2}$}\\
     & & \multicolumn{1}{c}{BMG} & \multicolumn{1}{c}{Coarse} & \multicolumn{1}{c}{Fine} & \multicolumn{1}{c}{Coarse} & \multicolumn{1}{c}{Fine} & \multicolumn{1}{c}{Coarse} & \multicolumn{1}{c}{Fine} & \multicolumn{1}{c}{Coarse} & \multicolumn{1}{c}{Fine}\\
     \hline \\[-1ex]
     \multicolumn{2}{c}{Post-Processing} & $159.46$  & $136.09$ & $135.99$ & $216.68$ & $181.52$ & $52.80$ & $68.47$ & $82.81$ & $73.16$\\ \\[-1ex]
     \multirow{ 4}{*}{0.0\%} & 20   & $303.85$    & $86.67$ &   $5.22$ & $41.13$ & $6.01$ & $23.52$ & $3.33$ & $47.39$ & $12.45$ \\
     & 40 & $182.53$ & $11.35$ & $2.66$ & $9.02$ & $4.42$ & $3.64$ & $2.06$ & $10.64$ & $6.51$\\
     & 100  & $159.46$  & $11.06$  & $1.94$ & $8.94$ & $6.26$ & $3.54$ & $2.71$ & $12.78$ & $10.51$\\
     & 160 & $156.81$ & $10.56$ & $2.01$ & $8.24$ & $4.29$ & $3.16$ & $1.80$ & $10.12$ & $5.18$\\
     \\[-1ex]
     \multirow{ 4}{*}{2.5\%} & 20 & $678.24$ & $228.64$ &  $181.25$ & $78.56$ & $65.56$ & $76.76$ & $76.40$ & $52.50$ & $65.45$\\
     & 40 & $464.45$ & $99.44$ & $77.20$ & $64.83$ & $65.34$ & $42.53$ & $41.36$ & $70.26$ & $68.60$\\
     & 100  & $357.86$ & $70.05$  & $58.74$ & $62.50$ & $63.86$ & $31.58$ & $48.82$ & $59.72$ & $65.80$\\
     & 160 & $307.16$ & $73.98$ & $64.01$ & $65.09$ & $63.26$ & $44.73$ & $49.60$ & $66.30$ & $61.65$\\
     \\[-1ex]
     \multirow{ 4}{*}{5.0\%} & 20 & $1138.96$ & $517.26$ & $331.74$ & $64.30$ & $126.04$ & $81.40$ & $85.87$ & $54.14$ & $71.59$\\
     & 40 & $917.07$ & $162.61$ & $242.86$ & $67.94$ & $109.05$ & $50.63$ & $53.81$ & $75.65$ & $76.71$\\
     & 100  & $700.81$  & $122.57$ & $97.83$ & $67.59$ & $83.56$ & $50.09$ & $36.48$ & $63.26$ & $64.16$\\
     & 160 & $571.20$ & $84.83$ & $84.81$ & $71.50$ & $66.38$ & $44.20$ & $49.96$ & $68.87$ & $63.63$\\
     \hline\hline
    \end{tabular}
    \label{tab:errors}
\end{table*}

The density of inputs, $(x,y,t)$, in space and time is of considerable interest because it affects the accuracy of reconstructions as well as the cost of training. There are two mechanisms by which accuracy is affected. First, residuals from \eqref{equ:physics residuals} should be zero throughout the measurement domain so adding collocation points effectively increases the amount of training data that is used to learn the flow. Second, the accuracy of the discrete projection model in \eqref{equ:model discrete} increases with the mesh resolution. Discrepancies between \eqref{equ:model continuous} and \eqref{equ:model discrete}, called \textit{model errors}, corrupt reconstructions in much the same way as measurement noise. Both mechanisms provide a strong incentive to maximize the number of inputs subject to constraints on one's computational resources.\par

Cai et al.~\cite{Cai2021a} found that significant artifacts arise when post-processing tomographic reconstructions with a PINN if the density of inputs falls below a critical threshold. We saw little change in the accuracy of post-processed fields upon increasing the node count from 1250 to 3600, indicating that our coarse grid of points already exceeded the minimum density threshold observed by Cai and his colleagues. By contrast, the same switch significantly improved the accuracy of direct reconstructions in key tests. For instance, while switching from our low-resolution grid to the high-res grid only reduced $\varepsilon_\mathrm{c}$ by 0.07\% in post-processing, concentration field errors in the direct reconstruction of 20 noise-free projections fell by over 90\% after changing grids. Crucially, $\varepsilon_\mathrm{c}$ decreased with increasing resolution in all direct tests that featured no noise or 2.5\% noise. However, accuracy gains associated with adding nodes diminished in the presence of 5\% noise, and even reversed in the 40-beam case (although that particular result was also adversely affected by our stopping criterion due to a sharp increase in $\varepsilon_\mathrm{c}$ following semi-convergence). Concentration field errors in the 160-beam test with 5\% noise were effectively invariant to our grids. This suggests that, given a sufficient number of projections, the resolution of reconstructions is limited by the intensity of noise rather than the density of $(x,y,t)$ inputs. Consequently, it is important to ensure the collection of clean data (or at least an ample amount of data) if one needs high-resolution reconstructions.\par

Directly reconstructed velocity and pressure fields were also enhanced through the use of a finer grid in our noise-free tests, whereas the corresponding fields obtained by post-processing $\mathbf{C}_\mathrm{est}$ were invariably degraded when we increased the number of inputs. By the same token, however, directly reconstructed velocity and pressure fields did not meaningfully improve upon switching to the high-resolution grid when the measurements were subject to noise. In fact, noise had a similar effect on direct estimates of $u$, $v$, and $p$ to the effect of reconstruction artifacts in $\mathbf{C}_\mathrm{est}$. This finding is corroborated by Fig.~\ref{fig:CUVP panel}, in which velocity and pressure fields recovered via post-processing or the direct reconstruction of noisy data bear little resemblance to the corresponding phantoms. Accordingly, errors in $u$, $v$, and $p$ were an order of magnitude lower in direct reconstructions of noise-free data than in all of our other tests. This suggests that the primary benefit of PINN-based reconstructions of real data may be limited to fields that are included in the measurement loss term.\par

Another important factor in tomographic imaging is the cost of high-speed lasers and cameras needed to acquire the projection data as well as constraints on the position of sensors (i.e., since they often take up considerable space). Therefore, it is important to employ a signal processing strategy that can maximally leverage sparse data. In our noise-free tests, direct reconstruction errors were more or less constant, but small gains were realized when we doubled the beam count from 20 to 40. Of course, the number of measurements needed to fully-resolve a flow is likely to be problem specific, with complicated flows requiring more data than simple ones. Regardless, our results demonstrate that the number of projections required to produce near-perfect reconstructions is quite low given high-fidelity measurements. When the measurements were subject to noise, however, adding more beams consistently improved the accuracy of concentration field estimates. This can be seen in Fig.~\ref{fig:noise test}: the 160-beam reconstruction closely matches the ground truth flow field. In general, adding projections helps to counteract noise (as expected).\par

Detrimental effects of noise on the accuracy of direct reconstructions are reported in sections~\ref{sec:demo:semi-convergence} and \ref{sec:demo:gamma}. However, we did not explore the role of noise in post-processing since \textit{idealized} post-processing already exhibited lower accuracy than direct reconstructions of less data, even when it had been corrupted by significant errors. Table~\ref{tab:errors} provides evidence that measurement noise can have severe repercussions in post-processing: BMG reconstructions of noise-free projections from 100 beams exhibited large artifacts that were not significantly mitigated by post-processing, moreover the accuracy of BMG estimates was substantially degraded by noise, as can be seen in the tabulated results. Our tests thus serve as a conservative estimate of the benefits of direct reconstruction compared to post-processing, thereby validating our claims to this effect made in section~\ref{sec:tomography:deep learning algorithms}.\par

\section{Bayesian Reconstruction}
\label{sec:Bayesian}
Measurement and model errors can affect one's interpretation of a signal so it is important to quantify the uncertainties associated with noise and the models used to process data. This is especially true in the context of ill-posed inverse problems, such as flow field tomography, due to the central role of prior information in regularizing estimates~\cite{Kaipio2006}. As with traditional iterative reconstruction algorithms, measurement errors give rise to semi-convergence when a PINN is used for tomographic imaging. While UQ is often done by propagating standard errors through the equations used for signal processing, either via a Taylor series expansion or Monte Carlo simulation~\cite{GUM1995}, these methods fail to account for the uncertainties associated with prior information. By contrast, the Bayesian framework for signal processing inherently accounts for prior information and it can accommodate model uncertainties~\cite{Kaipio2006,Kacker2003}. In this section, we show how a B-PINN (Bayesian PINN) can be used to conduct flow field tomography along with comprehensive UQ. In doing so, we also explicate the cause of semi-convergence that arises when using a C-PINN (i.e., a conventional or non-Bayesian PINN) to reconstruct noisy projections.\par

\subsection{Bayesian Inference with a PINN}
\label{sec:Bayesian:overview}
Bayesian inference is a statistical framework for parameter estimation that naturally facilitates UQ. In this framework, the data, model parameters, and unknown QoI are treated as random variables that are characterized by probability distributions or density functions (PDFs), denoted $f(\cdot)$. These functions code one's knowledge of the value taken by one or more variables, which may be updated in light of measurement information via Bayes' equation. When conducting Bayesian inference with a PINN, as proposed by Yang et al.~\cite{Yang2021}, we consider a vector of network parameters, $\boldsymbol\uptheta$, which contains all the weights and biases depicted in Fig.~\ref{fig:PINN architecture}, as well as a matrix of projection data, $\mathbf{B}$. It is also assumed that the Navier--Stokes and advection--diffusion residuals in $\mathbf{E}$ should be zero. Instead of generating a single PINN, the goal of Bayesian deep learning is to generate a distribution of networks and thereby a distribution of outputs, $(c,u,v,p)$. This \textit{posterior} distribution reflects uncertainties in the fields of interest associated with measurement noise and the network's architecture.\par

Bayesian deep learning consists in computing the posterior PDF of $\boldsymbol\uptheta$ given the measurements in $\mathbf{B}$, i.e., $f(\boldsymbol\uptheta|\mathbf{B})$, using Bayes' equation:
\begin{equation}
    f(\boldsymbol\uptheta|\mathbf{B}) = \frac{f(\mathbf{B}|\boldsymbol\uptheta) f_\mathrm{pr}(\boldsymbol\uptheta)}{f(\mathbf{B})} \propto f(\mathbf{B}|\boldsymbol\uptheta) f_\mathrm{pr}(\boldsymbol\uptheta).
    \label{equ:Bayes}
\end{equation}
Besides the posterior density, this expression contains three key PDFs:
\begin{enumerate}
    \item The likelihood, $f(\mathbf{B}|\boldsymbol\uptheta)$, describes the chance of observing projections, $\mathbf{B}$, of the flow field produced by a PINN with parameters $\boldsymbol\uptheta$. This chance is usually based on the distribution of measurement noise.
    
    \item The prior, $f_\mathrm{pr}(\boldsymbol\uptheta)$, encodes one's knowledge of the network parameters that is independent of the measurement information. In flow field tomography, the prior contains the physics residuals.
    
    \item The evidence, $f(\mathbf{B})$, quantifies the probability density associated with the data. Since the data are fixed in this context, the evidence is a constant that scales $f(\mathbf{B}|\boldsymbol\uptheta) f_\mathrm{pr}(\boldsymbol\uptheta)$ to ensure that the volume of $f(\boldsymbol\uptheta|\mathbf{B})$ is unity. That is,
    \begin{equation}
        f(\mathbf{B}) = \int f(\mathbf{B}|\boldsymbol\uptheta) f_\mathrm{pr}(\boldsymbol\uptheta) \,\mathrm{d}\boldsymbol\uptheta,
        \label{equ:evidence}
    \end{equation}
    which often needs to be calculated by an advanced numerical method due to the high dimension of $\boldsymbol\uptheta$.
\end{enumerate}
Our likelihood and prior functions are described in section~\ref{sec:Bayesian:likelihood and prior}, and section~\ref{sec:Bayesian:HMC} contains an overview of the sampling procedure used to approximate the posterior.\par

With the desired posterior density in hand, posterior distributions of the network's outputs are employed to visualize flow fields and quantify uncertainties thereof. We use the conditional mean (CM) as a representative point estimate,
\begin{equation}
    c_\mathrm{CM} = \int_{-\infty}^\infty c \,f(c|\mathbf{B}) \,\mathrm{d}c,
    \label{equ:posterior mean}
\end{equation}
although it is also common to employ the maximum a posteriori (MAP) estimate,
\begin{equation}
    c_\mathrm{MAP} = \mathrm{arg}\,\underset{c}{\mathrm{max}} [f(c|\mathbf{B})].
    \label{equ:MAP estimate}
\end{equation}
In these expressions, $f(c|\mathbf{B})$ is shorthand for the marginal PDF of $c$ at some location, $(x,y,t)$, outputted by the distribution of networks that corresponds to $f(\boldsymbol\uptheta|\mathbf{B})$; we also use $f(\mathbf{c}|\mathbf{B})$ and $f(\mathbf{C}|\mathbf{B})$ to indicate the posterior density of a vector of concentration data (e.g., a 2D reconstruction) or a sequence of such estimates, respectively. Uncertainties are presented in terms of an equal tailed $X$\% credible interval (CI), which satisfies
\begin{equation}
    \int_\alpha^\beta f(c|\mathbf{B}) \,\mathrm{d}c = \frac{X}{100},
    \label{equ:credible interval}
\end{equation}
where
\begin{equation}
    \int_{-\infty}^\alpha f(c|\mathbf{B}) \,\mathrm{d}c = \int_\beta^\infty f(c|\mathbf{B}) \,\mathrm{d}c
    \label{equ:equal tails}
\end{equation}
and $\Delta c_X = \beta - \alpha$ is the width of the interval. We use 95\% CIs in this paper.\par

\subsection{Likelihood and Prior PDFs}
\label{sec:Bayesian:likelihood and prior}
Equation~\eqref{equ:model continuous} relates a known concentration field to its projections, and \eqref{equ:model discrete} approximates \eqref{equ:model continuous} with a high degree of fidelity when the grid has sufficient spatial resolution. Therefore, provided that one has an adequate grid, discrepancies between the experimental measurements, $\mathbf{B}$, and projections of the \textit{true} flow field, $\mathbf{AC}_\mathrm{exact}$, are dominated by noise. In flow field tomography, the likelihood quantifies the chance of observing $\mathbf{B}$ for a given value of the unknown parameter vector, $\boldsymbol\uptheta$, which corresponds to a set of concentration fields outputted by the PINN, i.e., $\mathbf{C}(\boldsymbol\uptheta)$. This chance depends on the distribution of noise, which is often well modeled by a set of IID Gaussian random variables. The likelihood PDF of this error model is
\begin{equation}
    f(\mathbf{B}|\boldsymbol\uptheta) \propto \exp\mathopen{}\left[-\frac{\left\lVert\mathbf{B} - \mathbf{AC}(\boldsymbol\uptheta)\right\rVert_\mathrm{F}^2} {2\sigma_\mathrm{meas}^2}\right],
    \label{equ:likelihood PDF}
\end{equation}
where $\sigma_\mathrm{meas}$ is the standard deviation of the noise.\par

Substantive prior information about the network parameters is derived from the Navier--Stokes and advection--diffusion equations. That is, a PINN with parameters $\boldsymbol\uptheta$ should output flow fields that satisfy \eqref{equ:physics residuals}. The true distribution of physics residuals is unknown; hence, a Gaussian PDF is a natural choice because it is the maximum entropy distribution subject to a known mean and variance. In this case, the residuals should have zero mean so the prior density is
\begin{equation}
    f_\mathrm{pr}(\boldsymbol\uptheta) \propto \exp\mathopen{}\left[-\frac{\left\lVert \mathbf{E}(\boldsymbol\uptheta) \right\rVert_\mathrm{F}^2} {2\sigma_\mathrm{phys}^2}\right],
    \label{equ:prior PDF}
\end{equation}
where $\sigma_\mathrm{phys}$ is the standard deviation of the residuals. Selection of this parameter is a critical consideration that we discuss in section~\ref{sec:Bayesian:comparison}.

Depending on the network architecture, it may be necessary to modify \eqref{equ:prior PDF} with a cut-off or fade-out beyond some value of $\boldsymbol\uptheta$ since there exists an infinite set of flow fields that satisfy \eqref{equ:physics residuals}. Indeed, it is common to place a Gaussian prior on $\boldsymbol\uptheta$~\cite{Yang2021,Neal2012}. We also note that, while our prior does not contain any flow-specific information, there are advanced techniques for learning a better prior when additional, problem specific data are available~\cite{Meng2021a,Meng2021b}.\par

\subsection{Approximating the Posterior}
\label{sec:Bayesian:HMC}
As is almost always the case in Bayesian deep learning, there is no closed-form expression of the posterior distribution obtained by substituting our likelihood and prior PDFs into Bayes' equation, let alone the posteriors of $c$, $u$, $v$, and $p$, which are of primary interest in flow field tomography. For this reason, $f(\boldsymbol\uptheta|\mathbf{B})$ is usually approximated using a Markov chain Monte Carlo (MCMC) method or variational inference (VI)~\cite{Neal2012}. MCMC algorithms generate samples of the QoI that collectively behave as if drawn from the posterior as the chain grows longer. However, the network contains a large number of unknown QoI and the density of $f(\boldsymbol\uptheta|\mathbf{B})$ is almost always concentrated about a thin manifold. Consequently, common MCMC techniques fail due to the low probability of randomly approaching the dominant modes of $f(\boldsymbol\uptheta|\mathbf{B})$ in a high-dimensional space or the immense cost of looping through individual dimensions. Therefore, Bayesian neural nets are usually sampled using an MCMC variant called Hamiltonian Monte Carlo (HMC), which was devised to facilitate sampling in high-dimensions~\cite{Betancourt2017}. Yang et al.~\cite{Yang2021} computed the posterior PDF of several B-PINNs by HMC and VI. They found that HMC performed significantly better than VI so we utilized the Hamiltonian approach to sample our B-PINN posteriors, accordingly.\par

Hamiltonian Monte Carlo essentially replaces the random steps in a basic Metropolis--Hastings MCMC algorithm with a random \textit{trajectory} that is inspired by physics. Specifically, a hypothetical particle at some position, $\boldsymbol\uptheta$, in the probability space of interest is assigned a momentum, $\boldsymbol\uprho$, which has the same dimension as $\boldsymbol\uptheta$. The particle has kinetic energy ($V$) that depends on its momentum,
\begin{equation}
    V(\boldsymbol\uprho) = \frac{1}{2}\boldsymbol\uprho^\mathrm{T} \mathbf{M}^{-1} \boldsymbol\uprho,
    \label{equ:kinetic energy}
\end{equation}
where $\mathbf{M}$ is a ``mass matrix''; note that \eqref{equ:kinetic energy} mimics the kinetic energy formula from classical physics. Potential energy ($U$) of the particle is set to the negative log posterior,
\begin{align}
    U(\boldsymbol\uptheta) &= -\log{\left[f(\boldsymbol\uptheta|\mathbf{B})\right]} \nonumber\\
    &= \frac{1}{2}\left(\sigma_\mathrm{meas}^{-2} \mathcal{L}_\mathrm{meas} + \sigma_\mathrm{phys}^{-2} \mathcal{L}_\mathrm{phys}\right) + Z,
    \label{equ:potential energy}
\end{align}
which is found by plugging \eqref{equ:likelihood PDF} and \eqref{equ:prior PDF} into \eqref{equ:Bayes} (we do not specify $Z$ since it is a constant and thus does not affect the path). This function yields a high potential energy in regions of low density and vice versa such that the particle gravitates towards the dense modes of the posterior. Summing $U$ and $V$ results in the so-called Hamiltonian,
\begin{equation}
    H(\boldsymbol\uptheta,\boldsymbol\uprho) = U(\boldsymbol\uptheta) + V(\boldsymbol\uprho),
    \label{equ:Hamiltonian}
\end{equation}
which represents the total energy of a particle at locations in \textit{phase space}, i.e., at joint values of position and momentum, $(\boldsymbol\uptheta, \boldsymbol\uprho)$. Frictionless motion of the particle is governed by Hamiltonian mechanics, in which $\boldsymbol\uptheta$ and $\boldsymbol\uprho$ evolve according to
\begin{subequations}
\begin{align}
    \frac{\mathrm{d}\boldsymbol\uptheta}{\mathrm{d}t} &= \frac{\partial H}{\partial \boldsymbol\uprho} = \mathbf{M}^{-1}\boldsymbol\uprho \quad\text{and} \label{equ:Hamiltonian mechanics:position}\\
    \frac{\mathrm{d}\boldsymbol\uprho}{\mathrm{d}t} &= -\frac{\partial H}{\partial \boldsymbol\uptheta} = -\nabla U(\boldsymbol\uptheta). \label{equ:Hamiltonian mechanics:momentum}
\end{align}
\label{equ:Hamiltonian mechanics}%
\end{subequations}
Total energy is conserved by \eqref{equ:Hamiltonian mechanics}, i.e., $H$ is constant although $U$ and $V$ are typically in flux. This property can be exploited to sample the posterior PDF.\par

In HMC, a Markov chain of $\boldsymbol\uptheta$ is generated by drawing from a joint distribution of the QoI and momentum and then marginalizing the latter. To facilitate this, the momentum vector is assigned a centered Gaussian distribution with covariance $\mathbf{M}$, which is easy to sample. The corresponding momentum PDF is
\begin{equation}
    f(\boldsymbol\uprho) \propto \exp{\left[-V(\boldsymbol\uprho)\right]}.
    \label{equ:momentum PDF}
\end{equation}
Conveniently, the PDF of the particle's position is
\begin{equation}
    f(\boldsymbol\uptheta) \propto \exp{\left[-U(\boldsymbol\uptheta)\right]} = f(\boldsymbol\uptheta|\mathbf{B}),
    \label{equ:position PDF}
\end{equation}
and the joint density of $\boldsymbol\uptheta$ and $\boldsymbol\uprho$ is
\begin{equation}
    f(\boldsymbol\uptheta,\boldsymbol\uprho) \propto \exp{\left[-U(\boldsymbol\uptheta) - V(\boldsymbol\uprho)\right]} = \exp{\left[-H(\boldsymbol\uptheta,\boldsymbol\uprho)\right]}.
    \label{equ:joint PDF}
\end{equation}
Given a random momentum drawn from $f(\boldsymbol\uprho)$, new values of $\boldsymbol\uptheta$ and $\boldsymbol\uprho$ can be obtained without changing $f(\boldsymbol\uptheta, \boldsymbol\uprho)$ by traversing the phase space according to \eqref{equ:Hamiltonian mechanics} for a set duration. Repeating this procedure with successive draws of $\boldsymbol\uprho$ produces a chain of $(\boldsymbol\uptheta, \boldsymbol\uprho)$ samples that are proportional to $f(\boldsymbol\uptheta, \boldsymbol\uprho)$~\cite{Betancourt2017}. The marginal distribution of $f(\boldsymbol\uptheta, \boldsymbol\uprho)$ is itself proportional to the desired posterior PDF,
\begin{equation}
    \int f(\boldsymbol\uptheta, \boldsymbol\uprho) \,\mathrm{d}\boldsymbol\uprho \propto \exp{\left[-U(\boldsymbol\uptheta)\right]} = f(\boldsymbol\uptheta|\mathbf{B}),
    \label{equ:marginal Hamiltonian}
\end{equation}
per \eqref{equ:position PDF} and \eqref{equ:joint PDF}. In other words, the target Markov chain is given by $\boldsymbol\uptheta$ components of the phase space samples. To summarize, HMC utilizes random draws of $\boldsymbol\uprho$, which are straightforward to generate, in conjunction with the geometry of $f(\boldsymbol\uptheta|\mathbf{B})$, via \eqref{equ:Hamiltonian mechanics}, to efficiently explore the posterior of $\boldsymbol\uptheta$ given $\mathbf{B}$.\par

As mentioned earlier, B-PINNs are specified by a distribution of $\boldsymbol\uptheta$ as opposed to a single vector of parameters. Sample network parameters obtained by HMC are representative of this distribution, and marginal PDFs of the outputs can be approximated by evaluating a PINN with each vector $\boldsymbol\uptheta$ at the desired input locations.\par

\subsection{B-PINN Architecture and Sampling}
\label{sec:Bayesian:architecture}
Several measures were taken to limit the dimension of $\boldsymbol\uptheta$ and cost of training/sampling when comparing C-PINNs to B-PINNs. For instance, we used our coarse grid instead of the fine grid, we restricted training to a seqeunce of ten timesteps out of 201, and each PINN comprised five hidden layers instead of ten (although we maintained our initial choice of 50 neurons per output variable). Moreover, while most PINNs take advantage of weight normalization to improve the speed and stability of training~\cite{Salimans2016}, we opted for consolidated weight vectors, thereby reducing the number of weight parameters by half (we also verified that our C-PINN reconstructions were unaffected by this change). Of course, 3D flow fields will require a much larger probability space, but we leave implementation of a B-PINN for 3D flow field tomography as a topic for future research.\par

Conveniently, the gradient of $U$ in \eqref{equ:Hamiltonian mechanics:momentum} can be calculated by AD, but numerical integration of \eqref{equ:Hamiltonian mechanics} requires some extra scrutiny. The most common integration scheme in HMC is the leapfrog method~\cite{Betancourt2017}. The open source Python package hamiltorch~\cite{Cobb2020} utilizes PyTorch~\cite{Paszke2019} to conduct HMC sampling of a Bayesian neural network via leapfrog integration, and we implemented our B-PINNs using this framework. However, integration of the Hamiltonian using this package requires a user-specified step size and path length, both of which have a controlling influence on the accuracy and efficiency of HMC. Short steps and/or long path lengths are a waste of computational effort. By contrast, large steps are prone to numerical error and short paths result in diffusive motion (much like random walk MCMC with small steps). These outcomes are undesirable so either the step size and path length must be carefully tuned or else an adaptive HMC method should be employed, e.g., \cite{Hoffman2014}.  The mass matrix can also be tuned to introduce correlations between various components of $\boldsymbol\uprho$, but doing so effectively requires considerable insight about the morphology of the posterior PDF, which is unavailable in this application~\cite{Neal2012}. We conducted extensive parameter sweeps to identify an appropriate step size, path length, and mass matrix scale, but we will explore adaptive techniques in subsequent efforts. Lastly, some degree of numerical error is unavoidable, causing $H$ to fluctuate, which can lead to biased samples. This bias is counteracted with a modified Metropolis--Hastings acceptance criterion, as described by Betancourt~\cite{Betancourt2017}.\par

We set the mass matrix to be the identity matrix ($\mathbf{M} = \mathbf{I}$), the integration timestep to be $\epsilon = 5\times10^{-5}$, and the path length to be $50\epsilon$. Samples generated using these parameters exhibited desirable properties including a short correlation length scale and an adequate acceptance rate of roughly 44\%. In order to reduce the duration of burn-in, we ``hot-started'' the Markov chain at a $\boldsymbol\uptheta$ position obtained from a pre-trained C-PINN; burn-in lasted for approximately 600 samples, but we dropped the first 2000 to provide a factor of safety. The total chain length was 6000 in each case. On average, sampling in this manner took 44 computing hours on the same hardware described above. The average evaluation time for a single timestep was 74~milliseconds for a C-PINN and 448~seconds for a B-PINN. This is because the B-PINN comprises many networks to approximate the posterior distribution, which must be individually evaluated at each input coordinate.\par

\subsection{C-PINN Reconstruction vs. B-PINN Reconstruction}
\label{sec:Bayesian:comparison}
Comparative testing of C-PINNs and B-PINNs for flow field tomography commenced with two data sets consisting of clean measurements and noisy (2.5\%) measurements from 100 beams. We set the width of our likelihood to a nominal value in the noise-free tests, i.e., $\sigma_\mathrm{meas} = 0.005\cdot\mathrm{max}(\mathbf{B}_\mathrm{exact})$, and we used the actual value of $\sigma_\mathrm{meas}$ in our noisy tests. C-PINN reconstructions were conducted using $\gamma = 10^{-3}$, to start, which was found to be optimal in section~\ref{sec:demo:gamma}, and B-PINN reconstructions were carried out with a commensurate value of $\sigma_\mathrm{phys}$. At this point, it should be noted that the posterior produced by \eqref{equ:likelihood PDF} and \eqref{equ:prior PDF} is directly related to the total loss in \eqref{equ:total loss:measurement} used for direct reconstruction, as can be seen in the negative log posterior in \eqref{equ:potential energy}. In fact, maximizing $f(\boldsymbol\uptheta|\mathbf{B})$ is equivalent to minimizing $\mathcal{L}_\mathrm{total}$ when
\begin{equation}
    \sigma_\mathrm{phys} = \sqrt{\gamma} \,\sigma_\mathrm{meas},
    \label{equ:gamma to sigma}
\end{equation}
meaning that C-PINN reconstructions can be interpreted as MAP estimates. We thus assigned $\sigma_\mathrm{phys}$ using \eqref{equ:gamma to sigma} to synchronize the two methods.\par

\begin{figure*}[t]
    \centering
    \includegraphics[width=0.95\textwidth]{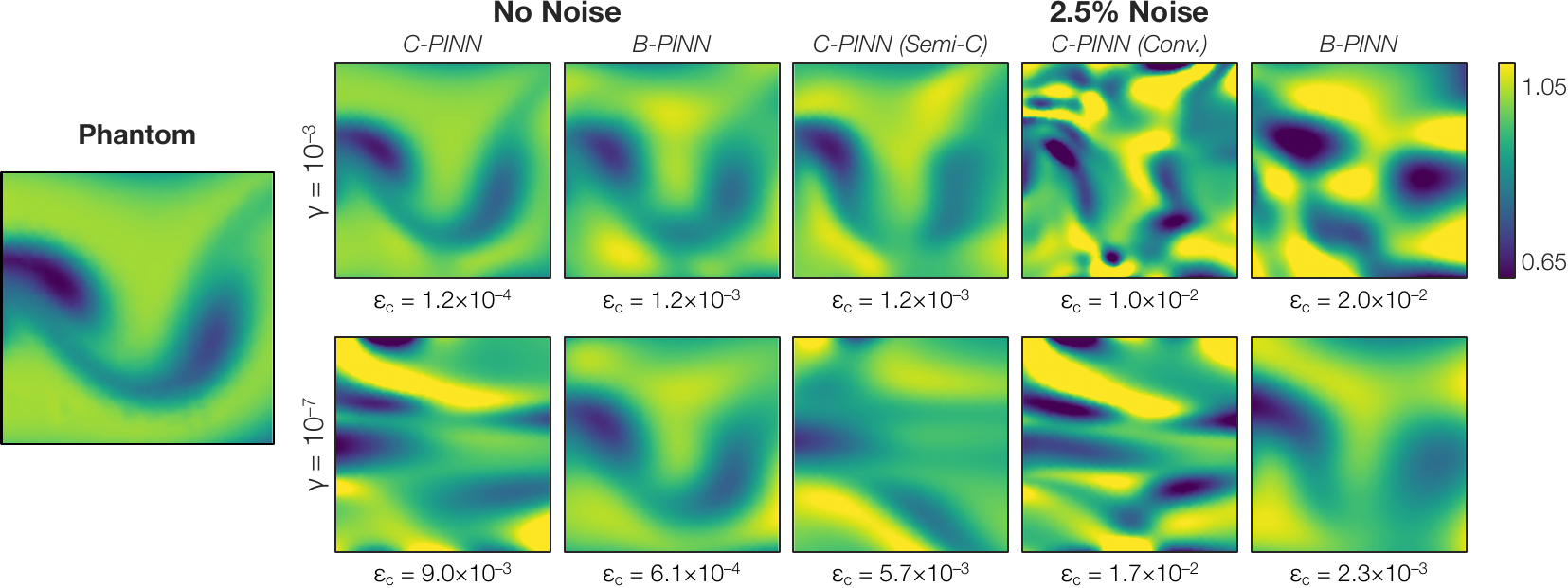}
    \caption{Reconstructions of clean and noisy data using C-PINNs and B-PINNs. The top row depicts reconstructions using values of $\gamma$ and $\sigma_\mathrm{phys}$ that were optimized for conventional reconstruction whereas the bottom row contains reconstructions using parameters optimized for Bayesian estimation. Increasing the weight of $\mathcal{L}_\mathrm{phys}$ should improve the accuracy of reconstructions, especially in the context of noisy data. However, this is only true of B-PINN estimates, which indicates that $f(\boldsymbol\uptheta|\mathbf{B})$ (and thereby $\mathcal{L}_\mathrm{total}$) contains one or more maxima (minima) that does not reflect the dominant posterior mode(s).}
    \label{fig:Bayesian panel}
\end{figure*}

Bayesian reconstructions yield PDFs of $(c, u, v, p)$ at each input coordinate, instead of a point estimate as in non-Bayesian algorithms. Hence, it is useful to plot the CM of these fields such that Bayesian estimates can be visually compared to conventional reconstructions. The top row of Fig.~\ref{fig:Bayesian panel} contains sample concentration fields produced by C-PINNs and B-PINNs using $\gamma = 10^{-3}$. Conventional estimates exhibited greater quantitative and qualitative accuracy than Bayesian estimates in our first set of noise-free tests. This difference is sensible since the Bayesian procedure assumed uncertainty in the measurements, despite the absence of noise, and measurement residuals had a significant effect at $\gamma = 10^{-3}$, as illustrated by the relative magnitude of $\mathcal{L}_\mathrm{meas}$ and $\mathcal{L}_\mathrm{phys}$ in Fig.~\ref{fig:training}. C-PINN estimates obtained using our stopping criterion performed well in our noisy tests when $\gamma$ was set to $10^{-3}$, whereas the fully-converged estimate and the Bayesian CM were dominated by artifacts. Per the discussion in section~\ref{sec:demo:semi-convergence}, this behavior simply reflects the information encoded in the total loss term or, equivalently, in the likelihood and prior PDFs.\par

Presumably, true flow fields will satisfy the Navier--Stokes and advection--diffusion equations. Therefore, so long as the network is deep and wide enough to approximate the flow and there are sufficient inputs and measurements for training, then it should be beneficial to set $\sigma_\mathrm{phys}$ to a much smaller value than $\sigma_\mathrm{meas}$, i.e., because we have a high degree of confidence in the governing physics. We thus conducted another set of comparisons using $\gamma = 10^{-7}$ and the corresponding value of $\sigma_\mathrm{phys}$. Concentration field estimates from these tests are shown in the bottom row of Fig.~\ref{fig:Bayesian panel}. All the flow fields outputted by a C-PINN that was trained using $\gamma = 10^{-7}$ were highly inaccurate, which we anticipated based on the weight parameter tests in section~\ref{sec:demo:gamma}. However, B-PINN reconstructions dramatically improved after reducing $\gamma$ in both the noise-free and noisy tests.\par

The stark difference between C-PINN and B-PINN reconstructions at $\gamma = 10^{-3}$ and $10^{-7}$ speaks to the topology of $\mathcal{L}_\mathrm{total}$ and $f(\boldsymbol\uptheta|\mathbf{B})$. Bayesian estimates depicted in Fig.~\ref{fig:Bayesian panel} represent the mean output of a distribution of networks whereas the flow fields obtained from a fully-converged C-PINN are essentially MAP estimates. C-PINN reconstructions of noisy data, regularized by our stopping criterion, are more difficult to characterize in this way because they do not minimize an explicit objective function. While the accuracy of CM concentration fields improved after switching from $\gamma = 10^{-3}$ to $10^{-7}$, tantamount to boosting the physics-based prior, the C-PINN reconstruction (akin to a MAP estimate) got far worse, as expected from the $\gamma$ tests shown in Fig.~\ref{fig:gamma test}. This difference in behavior was especially striking in the noisy case since \textit{none} of the fully-converged C-PINNs produced acceptable reconstructions of noisy data in any of our tests, including the networks considered in section~\ref{sec:demo}. This strongly suggests the surface corresponding to $f(\boldsymbol\uptheta|\mathbf{B})$ contains one (or more) robust maxima that does not reflect the dominant posterior mode(s). As a corollary, the same is likely true of minima in $\mathcal{L}_\mathrm{total}$. This implication was corroborated by varying randomly-selected weights of the network and plotting $\mathcal{L}_\mathrm{total}$. Each 4D plot of $\mathcal{L}_\mathrm{total}$ using $\gamma = 10^{-3}$ was consistent with a convex function (exhibiting a locally positive semidefinite Hessian) whereas plots using $\gamma = 10^{-7}$ were not. Consequently, CM estimates from a B-PINN are more representative of the functional used to reconstruct the flow than point estimates from a C-PINN, thereby facilitating the use of a wider range of priors. We leveraged this flexibility to implement a more stringent physics prior; CM reconstructions produced by this prior conformed to the true flow field, unlike any of the MAP estimates.\par

\subsection{UQ}
\label{sec:Bayesian:UQ}
\begin{figure}[t]
    \centering
    \includegraphics[width=.4\textwidth]{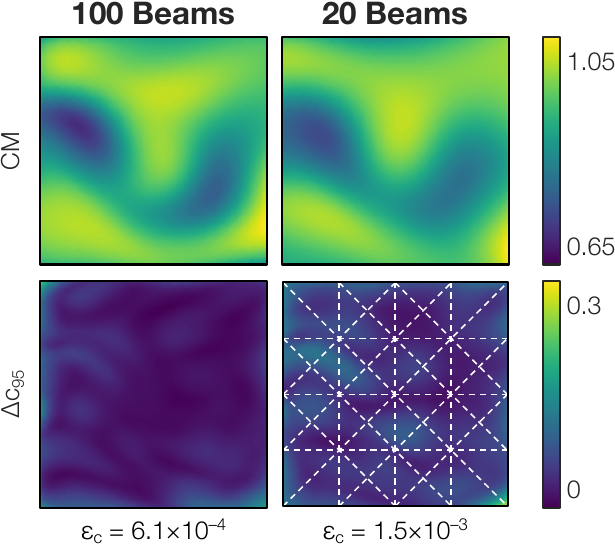}
    \caption{Conditional mean and CI (i.e., $\Delta c_{95}$) maps of the concentration field distributions outputted by a B-PINN. Reconstructions were conducted using clean projections from 100 beams and 20 beams. Beam paths are superimposed on the 20-beam CI to illustrate how regions of uncertainty are aligned with the gaps between beams.}
    \label{fig:posterior uncertainties}
\end{figure}

Bayesian results presented thus far consist of the mean concentration fields outputted by the chain of PINNs sampled from a posterior PDF. However, this distribution of networks conveys additional information about these flow fields, which can be leveraged to visualize and quantify the uncertainties produced by noise and model errors. Figure~\ref{fig:posterior uncertainties} contains plots of the CMs and CIs (i.e., $\Delta c_{95}$) obtained by reconstructing noise-free data from 100 beams and 20 beams, respectively, using the nominal value of $\sigma_\mathrm{meas}$ specified above and $\gamma = 10^{-7}$ in both tests. The CIs feature pockets and striations of elevated uncertainty that roughly align with the gaps between beams as well as a ring of uncertainty that wraps around the edge of the domain, where measurement information is in especially short supply. Beam paths are superimposed on the 20-beam CI map to emphasize these results. Furthermore, there was an overall increase in uncertainty in the 20-beam test, reflecting the relative lack of projection data. Results like these can be used to quickly assess the reliability of estimated flow structures by visual inspection.\par

\begin{figure*}[b]
    \centering
    \includegraphics[height=8cm]{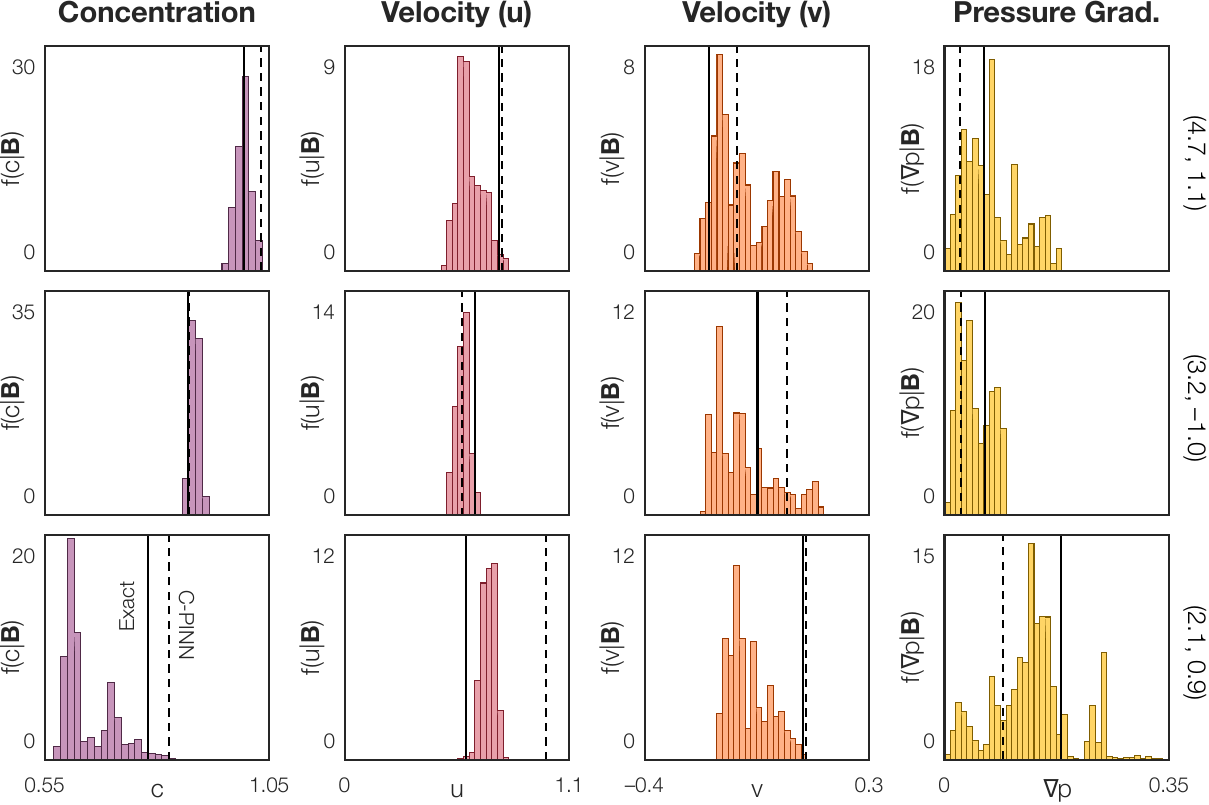}
    \caption{Marginalized posterior distributions of $(c,u,v,\nabla p)$ at three randomly selected points in the measurement domain, the $(x,y)$ location of each point is indicated at the right end of the row. Noisy (2.5\%) data from 100 beams was reconstructed using a C-PINN ($\gamma = 10^{-3}$ with iterative regularization) and a B-PINN ($\gamma = 10^{-7}$). Ground truth values are indicated with a solid black line, C-PINN estimates are plotted with a dashed line. Many of the distributions exhibit considerable non-Gaussianity.}
        \label{fig:posterior PDFs}
\end{figure*}

Still more information can be extracted from a B-PINN by plotting posterior PDFs of the flow fields. This is effectively demonstrated using the best-case C-PINN and B-PINN reconstructions of noisy data in Fig.~\ref{fig:Bayesian panel}, i.e., using $\gamma = 10^{-3}$ and iterative regularization to generate C-PINN estimates and $\gamma = 10^{-7}$ for the Bayesian reconstructions. Figure~\ref{fig:posterior PDFs} presents marginal posterior PDFs of the concentration, velocity, and pressure gradient fields at three randomly selected points in the domain. Ground truth values are indicated with a solid black line and C-PINN estimates are plotted with a dashed line. In some regions, the CM from the B-PINN was considerably closer to the ground truth value than the C-PINN estimate, whereas conventional PINN reconstructions were better in other regions. Treating CMs per se as a reconstruction leads to the conclusion that our B-PINN estimates were less accurate than our C-PINN estimates, which is apparent in Fig.~\ref{fig:Bayesian panel}. However, the PDFs in Fig.~\ref{fig:posterior PDFs} illustrate the wide range of $(c, u, v, \nabla p)$ values that were consistent with the data and physics prior throughout the domain. Notably, the ground truth values almost always lay in a region of high posterior density. By comparison, C-PINN results required iterative regularization and were less amenable to UQ for this reason. Moreover, Bayesian estimates were obtained using a much more intuitive functional, which was robust to noise, and the B-PINN procedure could likely be refined via improved sampling, using more timesteps, increasing the grid resolution, and so on. We thus conclude that the information produced by a B-PINN is more useful than point estimates from a C-PINN when using reconstructions for quantitative comparisons, e.g., to benchmark the results of a CFD simulation.\par

\section{Conclusions}
\label{sec:conclusion}
Physics-informed neural networks have been used to post-process tomographically-reconstructed flow fields in order to improve their accuracy and infer additional fields~\cite{Cai2021a}. By contrast, this work reports the first use of a PINN to directly reconstruct all the flow fields from a set of projection data by embedding the projection model into the PINN's data loss function. We also implemented a Bayesian PINN to directly reconstruct flow fields with built-in UQ. Several important conclusions can be drawn from this work.
\begin{enumerate}
    \item Direct reconstruction with a PINN was far more accurate than PINN-based post-processing of the reconstructions produced by a conventional algorithm. This remained true even when direct reconstructions were conducted with fewer, more noisy projections than the data set used for post-processing.
    
    \item Three regimes of training were identified in the direct reconstruction of noise-free projections. However, noise-affected data resulted in semi-convergence near the end of phase II. Training beyond this point resulted in overfitting to noise.
    
    \item We devised a stopping criterion, based on the transition from phase I to II of training, used to regularize direct reconstructions of noisy data. Our technique accurately approximated the point of semi-convergence.
    
    \item Bayesian PINNs were implemented to conduct UQ. By sampling the posterior instead of maximizing it, a B-PINN can produce CM estimates that better characterize the objective function than reconstructions from a C-PINN (akin to a MAP estimate). Bayesian learning thus facilitates the use of a wider range of objective functions than conventional training. We exploited this capability to implement a stricter physics prior, resulting in accurate reconstructions of noisy projections.
    
    \item Marginal posterior flow field PDFs produced by a B-PINN can be used for quantitative model comparison and validation purposes.
\end{enumerate}

\section*{Acknowledgments}
We thank Dr. Shengze Cai and Prof. George Em Karniadakis for their helpful comments.

\bibliographystyle{IEEEtran}

\end{document}